\documentclass[final,1p,times]{elsarticle}

\journal{Journal of Computational Physics}

\usepackage{mathrsfs,amsmath,amssymb,bm}
\usepackage{hyperref}
\usepackage{graphicx}
\usepackage{subfigure}
\usepackage{caption}
\usepackage{overpic}
\usepackage{epsfig}
\usepackage{xcolor}
\usepackage{indentfirst}
\usepackage{fancyhdr}
\usepackage{comment}
\usepackage{makecell, rotating}
\usepackage{algorithm} 
\usepackage{algorithmic} 
\usepackage{float}
\usepackage{fullpage} 

\usepackage{xkeyval}
\usepackage{todonotes}
\presetkeys{todonotes}{inline}{} 
\usepackage[draft]{changes}
\definechangesauthor[name={Huayi Wei}, color=red]{why}

\newcommand{\bx}{\mathbf{x}}

\newcommand{\calS}{\mathcal{S}}

\begin{document}

\title{
A finite element method of the self-consistent field theory
on \textcolor{blue}{general curved surfaces}}

\author{Huayi Wei}
\author{Ming Xu}
\author{Wei Si}
\author{Kai Jiang\corref{cor}}
\address{
 School of Mathematics and Computational Science, Xiangtan
 University, Xiangtan, Hunan, P.R. China, 411105,
 \\
Hunan Key Laboratory for Computation and Simulation in Science and Engineering
 }
 \cortext[cor]{Corresponding author. Email: kaijiang@xtu.edu.cn.}

\date{\today}


\begin{abstract}
	 Block copolymers provide a wonderful platform in studying
	 the soft condensed matter systems. Many fascinating ordered
	 structures have been discovered in bulk and confined
	 systems. Among various theories, the self-consistent field
	 theory (SCFT) has been proven to be a powerful tool for studying the
	 equilibrium ordered structures. Many numerical methods have
	 been developed to solve the SCFT model. However, most of
	 these focus on the bulk systems, and little work on the
	 confined systems, especially on general curved surfaces.
	 In this work, we developed a linear surface finite element
	 method, which has a rigorous mathematical theory to
	 guarantee numerical precsion, to 
	 study the self-assembled phases of block copolymers on 
	 general curved surfaces based on the SCFT.
	 Furthermore, to capture the consistent surface for a
	 given self-assembled pattern, an adaptive approach to
	 optimize the size of the general curved surface has been
	 proposed. To demonstrate the power of this approach, we
	 investigate the self-assembled patterns of diblock
	 copolymers on several distinct curved surfaces, including
	 five closed surfaces and an unclosed surface. Numerical results
	 illustrate the efficiency of the proposed method. The obtained
	 ordered structures are consistent with the previous results
	 on standard surfaces, such as sphere and torus. Certainly,
	 the proposed numerical framework has the capability of
	 studying the phase behaviors on general surfaces
	 precisely.
\end{abstract}

\maketitle

\section{Introduction}
\label{sec:intrd}

Geometry plays a key role in many scientific
fields\,\cite{nakahara2003topology}, including Hamiltonian mechanics with
constraints, general relativity, quantum mechanics, quantum field theory, the
arrangement of electrons on a sphere (Thomson problem), defect motion on a
curved surface, polymer field theory, generic reaction-diffusion model
occurring in chemistry and biology.  Recent years, the microphase separation of
block copolymer under various types of geometrical confinements both in the
bulk and on surfaces has been attracted tremendous attention.  Many novel
patterns of block copolymer emerge from rearrangements of traditional ordered
phases when adapting to these restrictions\,\cite{wu2004composite, xiang2005,
yu2006prl, charlotte2011interplay}.  The self-assembly of block copolymers
under geometrical constraint provides an efficient means on the nanometer scale
for such applications as the construction of high-capacity data storage
devices, waveguides, quantum dot arrays, dielectric mirrors, nanoporous
membranes, nanowires, and interference
lithography\,\cite{charlotte2011interplay, segalman2005patterning}. The
curvature effects on the self-assembled behavior of block copolymers
confined to a curved surface, however,  are still far from being fully
understood\,\cite{chantawansri2007, li2014self, li2006self}.  The main obstacle
may be that it still lacks a general theoretical framework which can properly
describe the microphase separation of block copolymers on general curved
surfaces.

Theoretically, the self-consistent field theory
(SCFT)\,\cite{fredrickson2006equilibrium} proposes a successful framework in
the study of the self-assembling of block copolymers.  It has predicted a
variety of ordered bulk phases that have been observed in
experiments, \textit{e.g.},
lamella, hexagonal cylinder, sphere and gyroid phases of diblock
copolymers\,\cite{matsen1994stable}, and many intricate ordered phases of
triblock copolymers\,\cite{jiang2015self}. Mathematically, the
SCFT model is a very complicated variational problem, possessing
many unsatisfactory features, such as saddle-point,
nonlinearity, multi-solutions, and multi-parameters.
Analytically solving this model goes beyond current technologies.
An alternating approach is the numerical technique.  

\textcolor{blue}{
The numerical methods for solving SCFT model mainly consists of four
components, namely screening
initial values\,\cite{xu2013strategy, jiang2010spectral,
jiang2013discovery}, solving time-dependent partial differential
equations (PDEs), evaluating (monomer) density operators,
and finding saddle-points \textit{via} iteration
methods\,\cite{chantawansri2007, ceniceros2004numerical, 
thompson2004improved}. }
In this paper, we focus on the spatial discretization of the 
time-dependent PDEs in the SCFT model. 
In flat spaces, many numerical approaches have been developed for  
spatial variables which can generally be divided into two
classes of discretization schemes. The first type is the
subspace discretization method which discretizes
equations in a constrained space based on specific properties.
For example, according to the crystallographic symmetry of a given
pattern in microphase-separated block copolymers, the PDEs in
SCFT model can be expanded in terms of a set of symmetric basis
functions. One of the typical representatives
is the symmetry-based Fourier method proposed by Masten and
Schick\,\cite{matsen1994stable}.  The second type is numerically
approximating PDEs in the whole space. These methods can be carried out
either in the real space\,\cite{drolet1999combinatorial} or in
the Fourier-space\,\cite{guo2008discovering}.  It has also been
demonstrated that the whole-space discretization methods are able
to discover new patterns\,\cite{drolet1999combinatorial,
guo2008discovering}. In recent years, an efficient Fourier
pseudospectral method has been introduced to solve PDEs in SCFT
model\,\cite{rasmussen2002improved,
cochran2006stability}.  It fully takes advantage of the best
performance of real space and Fourier-space and reduces
the computational complexity to $O(M\log M)$, with the number of
degrees $M$, based on the Fast Fourier Transformation (FFT). 
\textcolor{blue}{
Besides space discretization, many time discretization
schemes, such as the operator-splitting method, the
Crank-Nicokson scheme, and linear multi-step approaches, have been
used in the time
discretization\,\cite{fredrickson2006equilibrium}. 
The total computational complexity in solving time-dependent PDE
should consider both the time and spatial discretization schemes.
}

From flat spaces to curved surfaces, the Laplace operator in the PDEs becomes
the Laplace-Beltrami operator which is the heat kernel on the 
manifold. Numerical methods should be developed to discretize
the differential operator on a curved surface.
There are two approaches
to represent the curved surface. The first one is viewing the
curved surface as a level set through embedding it into a high
dimensional flat space.
In another viewpoint, the curved surface is represented by the inner
coordinates of the manifold. From the first viewpoint, the mask
method, which uses a large enough domain in high dimensional flat
space to cover the curved surface, can be chosen to attack 
curved surface problems\,\cite{osher2006level}. These methods
developed in the flat space can be used to solve the curved surface
problem. Then these results on a manifold can be
approximated by restricting the results of high dimensional flat space
on the curved surface. From another
viewpoint, one shall directly discretize the function defined on a curved surface. For
some special curved surfaces, such as the sphere surface, one can expand the
function defined on the sphere by the eigenfunctions,
\textit{i.e.}, the
spherical harmonic functions,  of the
Laplace-Beltrami operator\,\cite{chantawansri2007}.
For a general case, however, it is impossible to find out the
eigenfunctions of the Laplace-Beltrami operator.
It causes inconvenience in solving PDEs in the SCFT model on
general curved surfaces.  In 2014, Li \textit{el
al.}\,\cite{li2014self} mimicked the process of finite difference
method and developed an extended spherical ADI finite
difference\,\cite{li2006self} to solve SCFT model on a general
curved surface. However, the precision of this method can not be
guaranteed.

It is well-known that as an effective
numerical method for solving PDE, the finite element method plays an important
role in modern scientific and engineering computing.  Based on variational
principle, finite element method subdivides the definition domain into
small and simple regions, such as triangle, quadrilateral, tetrahedron,  or
hexahedron, \textit{etc.}, and then simple algebraic equations on each small region can
be created. At last these simple equations are assembled into a
large sparse algebraic system to approximate the PDE problem. The
advantage of the finite element method is that it can handle 
complex geometrical domains with complex boundaries.
Based on the similar principal, surface finite element method has
been developed to solve PDEs on a general
manifold\,\cite{Dziuk1988, Dziuk;Elliott2007, wei2010, Dziuk1991,
Dziuk2007, Demlow2009, Dziuk;Elliott2013}.  
The approach was firstly proposed by Dziuk to solve the
Laplace-Beltrami equation on arbitrary surfaces using linear
finite element\,\cite{Dziuk1988}. Then Dziuk and Elliott
\cite{Dziuk;Elliott2007} applied the linear surface finite
element to parabolic equations, and gave error
bounds in the case of semi-discretization in space for a fourth
order linear problem.  Our previous work \cite{wei2010}
generalized the superconvergence results and several gradient
recovery approaches of linear finite element methods from flat
spaces to general curved surfaces for the Laplace-Beltrami
equation with mildly structured triangular meshes.  In this work,
we will continue to extend the surface finite element method to
study the microphase separation of block
copolymer on general curved surface based on the SCFT model.

On the other hand, the surface size also affects self-assembling patterns.  For
example, the self-assembled periodic structures are affected by the
computational domain in flat space\,\cite{matsen1994stable}.  On the sphere
surface, the radius can affect the number of microdomains of a cylindrical
phase as well as its energy value, and phase behavior of lamellar phases.
Therefore, for a given self-assembled phase, there exists the
most appropriate size of curved surface.
To capture the optimal size, in this work, we will propose an
adaptive method to optimize the surface,
and further obtain accurate energy value for a given ordered structure.

The remaining portion is arranged as follows. In Sec.\,\ref{sec:pre}, we
introduce the preliminary knowledge of surface finite element method and
curved surfaces to be used in this paper.  In
Sec.\,\ref{sec:scft} we derive the SCFT model on the general curved
surface. The numerical algorithm of linear surface finite element
for solving SCFT model is given in Sec.\,\ref{sec:method}.
The efficiency of our method and the self-assembled structures of
diblock copolymers on several curved surfaces are presented in
Sec.\,\ref{sec:rslt}. In the Sec.\,\ref{sec:conclusion}, we will draw the
conclusion and outline the further work.

\section{Preliminaries}\label{sec:pre}

In this section, we firstly introduce some background knowledge about the
surface finite element method, and refer to \,\cite{Dziuk;Elliott2013} for
details. Then we present the surfaces used 
in the following calculations and the approach of how to generate
optimal meshes on these surfaces.

\subsection{Surface finite element method}

Let $\calS$ be a two-dimensional, compact, and $\mathcal C^2$-hypersurface in $\mathbb
R^3$.  Here we assume there exists a smooth level set function $d(\bx)$
that can describe the surface $\calS$ as  
\begin{equation*}
    \calS = \{\bx\in U ~|~ d(\bx) = 0\},
\end{equation*}
where $U$ is open subset of $\mathbb R^3$ in which $\nabla d(\bx) \not=0$, and
$d\in \mathcal C^2(U)$. We define the unit normal vector on
$\calS$ by 
\begin{equation*}
    \boldsymbol n = \frac{\nabla d(\bx)}{|\nabla d(\bx)|},
\end{equation*}
which means that the orientation of $\calS$ is fixed through normal
vector $\boldsymbol n$. For any $\bx \in U$, one can define a projection
\begin{equation*}
    \mathcal P(\bx) = \bx - d(\bx)\boldsymbol n,
\end{equation*}
such that $\mathcal P(\bx) \in \calS$. 

For $v(\bx)\in \mathcal C^1(\calS)$, since $\calS$ is $\mathcal C^2$, we can extent $v$ to $\mathcal C^1(U)$
and still denote the extension by $v$. The tangential gradient of
$v$ on $\calS$ can be written as 
\begin{equation*}
    \nabla_S v = \nabla v - (\nabla v\cdot\boldsymbol n)\boldsymbol n =
    \boldsymbol P(\nabla v),
\end{equation*}
where $\boldsymbol P(\bx) = (I - \boldsymbol n\boldsymbol n^t)(\bx)$ is the projection operator
to the tangent plane at a point $\bx\in \calS$. Notice that we use the extension of
$v$ to define the surface gradient. However, it can be shown that $\nabla_S v$ depends only
on the value of $v$ on $\calS$ but not on the
extension\,\cite{Demlow2009}. Namely, $\nabla_S$ is an intrinsic
operator. 

Similarly, for a vector field $\boldsymbol v \in (\mathcal
C^1(\calS))^3$, we can also extend it
to $(\mathcal C^1(U))^3$ and define the tangential divergence operator of
$\boldsymbol v$ on $\calS$ as
\begin{equation*}
    \nabla_S\cdot \boldsymbol v(\bx) = \nabla\cdot\boldsymbol v - \boldsymbol
    n^t\nabla \boldsymbol
    v\boldsymbol n,
\end{equation*}

The Laplace-Beltrami operator on $\calS$ reads as follows:
\begin{equation*}
\Delta_\calS v = \nabla_\calS\cdot(\nabla_\calS v) = \Delta v - (\nabla v\cdot\mathbf
n)(\nabla\cdot\boldsymbol n) - \boldsymbol n^t\nabla^2 v \boldsymbol n,
\end{equation*}
provided $v(\bx) \in \mathcal C^2(\calS)$ and $\nabla^2 v$ is the Hessian matrix of $v$ (suitably
extended as a $\mathcal C^2(U)$ matrix function). The Sobolev
spaces on surface $\calS$ can be defined as:
\begin{equation*}
    H^1(\calS) = \{v\in L^2(\calS)\, |\, \nabla_S v \in
	(L^2(\calS))^3 \}.
\end{equation*}

Let $\calS$ be approximated by a triangular mesh $\calS_h$ with node set $\mathcal
N_h = \{\bx_i\}_{i=1}^{N}$ and triangle set $\mathcal T_h = \{\tau_h\}$.  We
assume that these triangles are shape-regular and quasi-uniform of a
diameter $h$ and their vertices lie on $\calS$. For any $\tau_h\in \mathcal T_h$,
let $\boldsymbol n_h$ be the unit outward normal vector of $S_h$ on $\tau_h$.
Let $V_h$ be the continuous piecewise linear finite element space on $S_h$,
with linear Lagrangian basis functions $\{\varphi_i(\bx)\}_{i=1}^N$ defined
on $S_h$. $\varphi_i(\bx)$ is the piecewise linear function on each
triangle face $\tau_h$,
\begin{equation}\label{eq:varphi}
    \varphi_i(\bx_j) = 
    \begin{cases}
        1,&\text{ if } i=j,\\
        0,&\text{ if } i\not=j.
    \end{cases} 
\end{equation}
For $v_h\in \mathcal C(S_h)$ and $v_h|_{\tau_h}\in \mathcal C^1(\tau_h)$, we have 
\begin{equation*}
    \nabla_{S_h} v_h|_{\tau_h} := \nabla v_h - (\nabla v_h\cdot\boldsymbol
    n_h)\boldsymbol n_h = (\boldsymbol I - \boldsymbol n_h \boldsymbol
    n_h^t)\nabla v_h = \boldsymbol P_h\nabla v_h.
\end{equation*}

\subsection{Surface representation and mesh generation}

In this subsection, we will discuss how to generate high-quality 
finite element meshes on surfaces.
A high-quality mesh on a surface means:
the triangle elements are all almost equilateral and more mesh
nodes located at the place with large curvature\,\cite{wei2010}.
In this work, five closed curved surfaces including sphere,
torus, double torus, heart, and orthocircle surfaces, and an unclosed 
parabolic surface are considered.
In general, there are three ways to represent a continuous surface in $\mathbb R^3$:
explicitness, implicitness, and parametrization. 

\begin{figure}[H]
\setlength{\captionmargin}{2pt}
\centering
\subfigure[]{\includegraphics[scale=0.35]{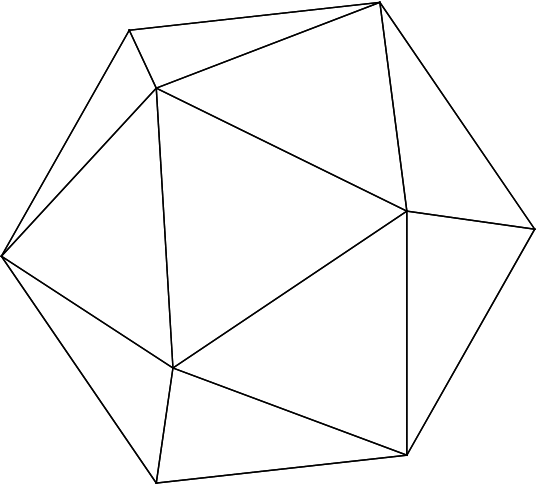}}
\hspace{0.5cm}
\subfigure[]{\includegraphics[scale=0.40]{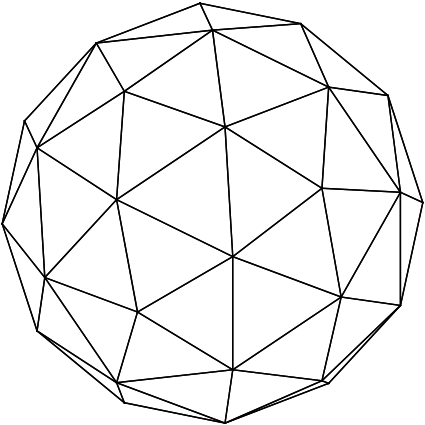}}
\hspace{0.5cm}
\subfigure[]{\includegraphics[scale=0.38]{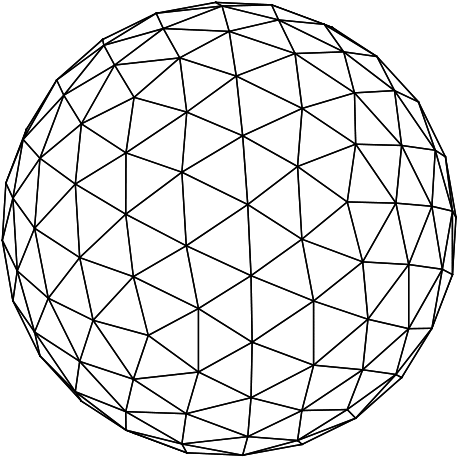}}
\hspace{0.5cm}
\subfigure[]{\includegraphics[scale=0.38]{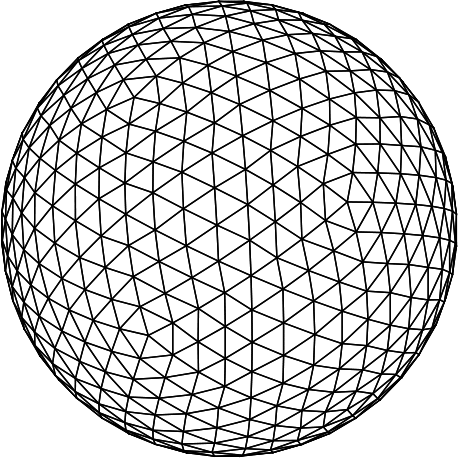}}
\caption{The uniform refinement icosahedron for generation triangular meshes on the unit sphere.}
\label{fig:mesh-sphere}
\end{figure}

\begin{description}
    \item[Sphere surface] 
\end{description}
Any point on the unit sphere surface satisfies the 
implicit equation $ \sqrt{x^2 + y^2 + z^2} - 1 = 0$.
A good triangular mesh on sphere can be easily
generated by uniformly refining an icosahedron and projecting the
new mesh nodes, which are the old mesh edge center, onto the sphere.
The refinement procedure is illustrated in
Fig.\,\ref{fig:mesh-sphere}. 

\begin{figure}[H]
\setlength{\captionmargin}{2pt}
\centering
\includegraphics[scale=0.45]{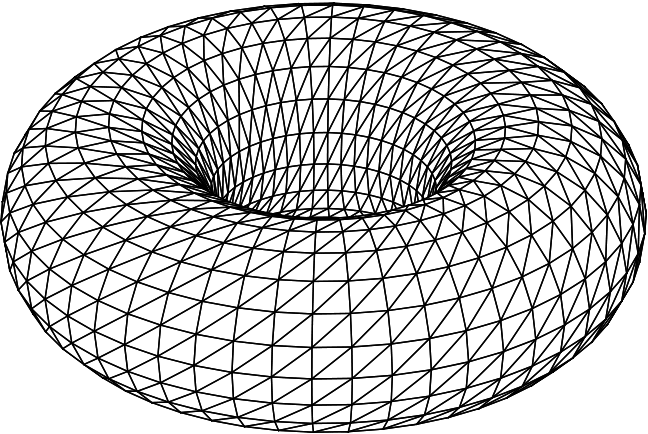}
\caption{The triangular mesh on torus surface.}
\label{fig:mesh-torus}
\end{figure}

\begin{description}
    \item[Torus surface] 
\end{description}
The second example is the torus which can be 
defined parametrically by
\begin{align*}
    x(\theta, \phi) &= (R + r\cos\theta)\cos\phi,\\
    y(\theta, \phi) &= (R + r\cos\theta)\sin\phi,\\
    z(\theta, \phi) &= r\sin\theta,
\end{align*}
where $\theta, \phi\in[0,2\pi]$, $R$ is the major radius which is
the distance from the center of the tube to the center of the torus, $r$ is the
minor radius which is the radius of the tube.
The ratio of $r/R$ is the so-called "aspect ratio". 
With the above parameteric representation, the
triangular mesh on torus can be generated by mapping a
triangular mesh on the square domain $[0, 2\pi]^2$, see
Fig.\,\ref{fig:mesh-torus}. 

\begin{figure}[H]
\setlength{\captionmargin}{2pt}
\centering
\subfigure[Double torus
surface.]{\includegraphics[scale=0.5]{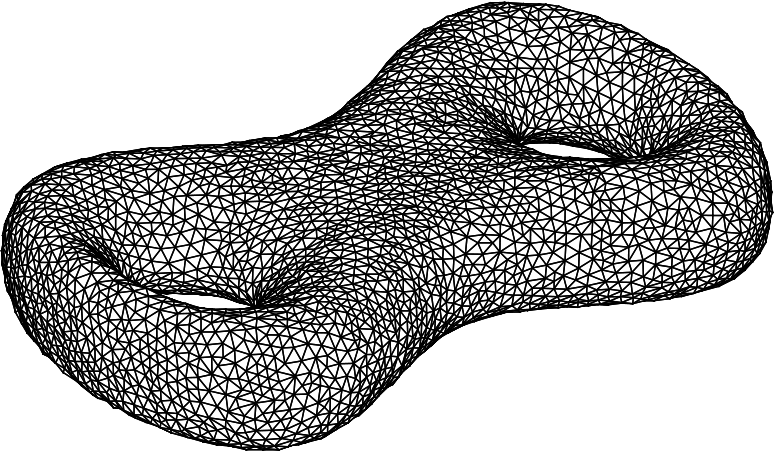}}
    \hspace{1cm}    
\subfigure[Heart surface.]{\includegraphics[scale=0.55]{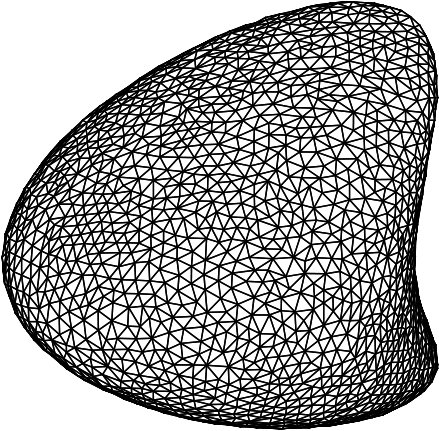}}
    \hspace{1cm}
\subfigure[Orthocircle surface.]{\includegraphics[scale=0.483]{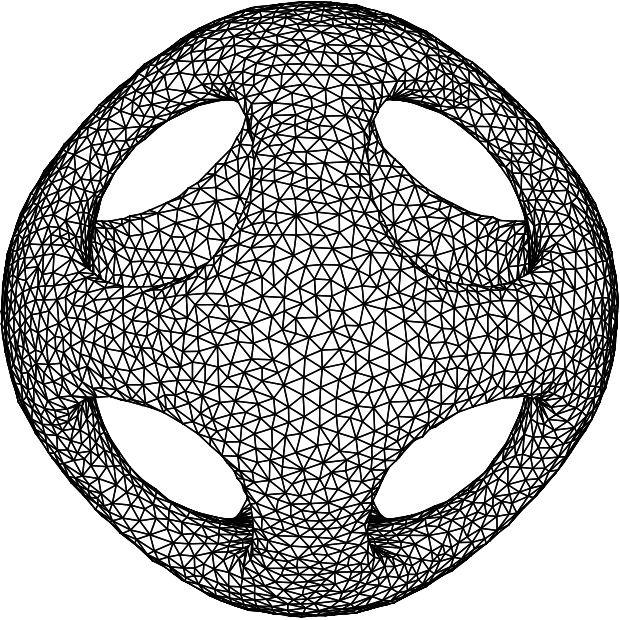}}
\caption{The initial triangular meshes on general surfaces
generated by 3D Surface Mesh Generation package, CGAL.}
\label{fig:mesh-CGAL-init}
\end{figure}

\begin{figure}[H]
\setlength{\captionmargin}{2pt}
\centering
\subfigure[Double torus
surface.]{\includegraphics[scale=0.5]{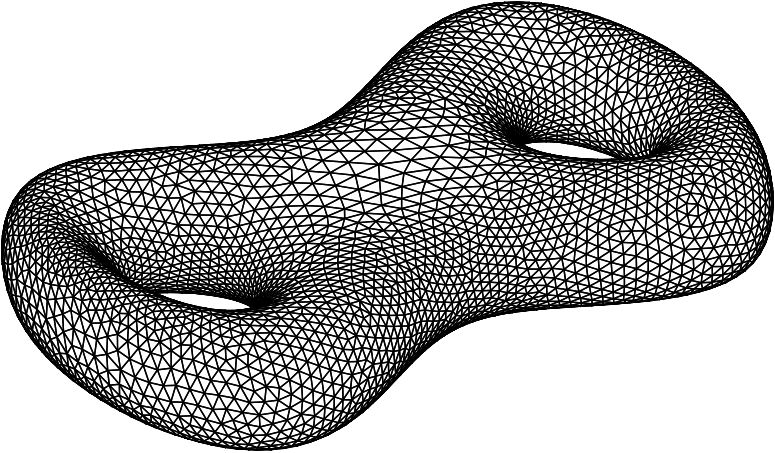}}
    \hspace{1cm}    
\subfigure[Heart surface.]{\includegraphics[scale=0.55]{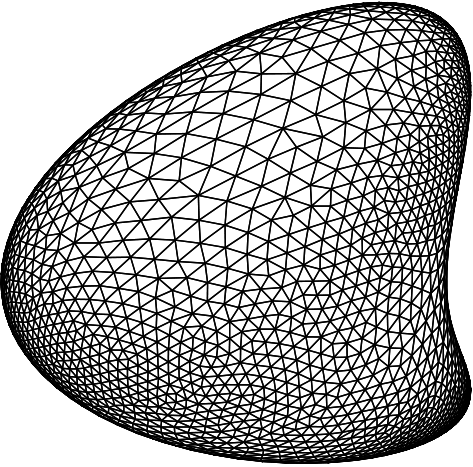}}
    \hspace{1cm}
\subfigure[Orthocircle
surface.]{\includegraphics[scale=0.483]{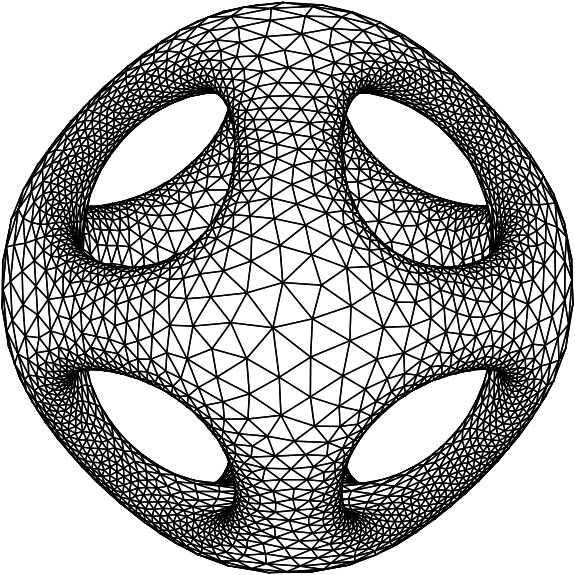}}
\caption{The optimized triangular meshes on general surfaces
through the CVT technique.}
\label{fig:mesh-CGAL}
\end{figure}

The implicit expressions of double torus, heart, and
orthocircle are given as follows:
\begin{description}
    \item[Double torus surface] 
		\begin{align}
        x^2(x^2-1)[x^2(x^2-1) + 2y^2] + y^4+ z^2 -0.04 =0.
			\label{fig:mesh:doubletorus}
		\end{align}
    \item[Heart surface] 
		\begin{align}
		(x-z^2)^2 +y^2+z^2-1=0.
			\label{fig:mesh:heart}
		\end{align}
    \item[Orthocircle surface]
		\begin{align}
        [(x^2+y^2-1)^2+z^2][(y^2+z^2-1)^2+x^2][(z^2+x^2-1)^2+y^2]-0.075^2[1+3(x^2+y^2+z^2)]=0.
			\label{fig:mesh:orthocircle}
		\end{align}
\end{description}
For these complicated surfaces,
it is difficult to generate a high quality surface triangluar mesh
as done for sphere or torus. To get the high quality
mesh, we firstly use the 3D Surface Mesh Generation package in
CGAL \cite{rieau2018} to generate an initial triangluar mesh
(see Fig.\,\ref{fig:mesh-CGAL-init}), 
then further optimized these initial mesh through
the Centroidal Voronoi Tessellation (CVT) technique
\cite{wei2012} (see Fig.\,\ref{fig:mesh-CGAL}). 

\begin{description}
    \item[Parabolic surface] 
\end{description}

\begin{figure}[H]
\setlength{\captionmargin}{2pt}
\centering
\subfigure[]{\includegraphics[scale=0.35]{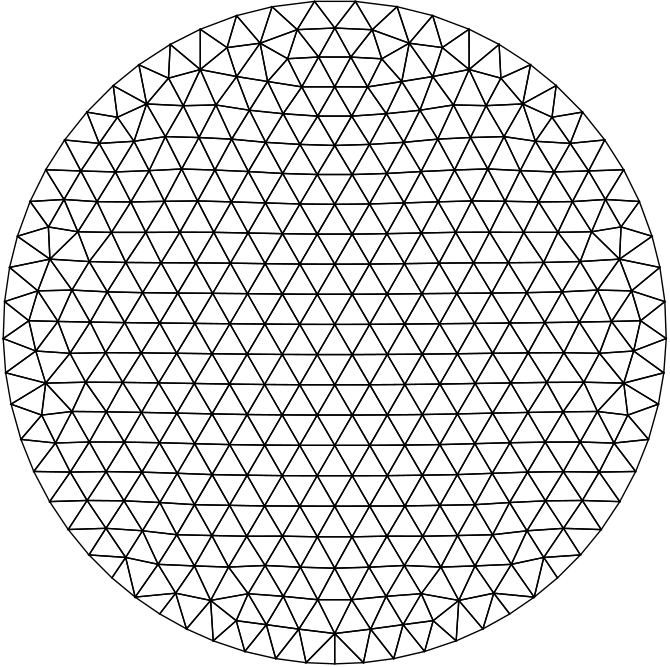}}
\hspace{1cm}
\subfigure[]{\includegraphics[scale=0.45]{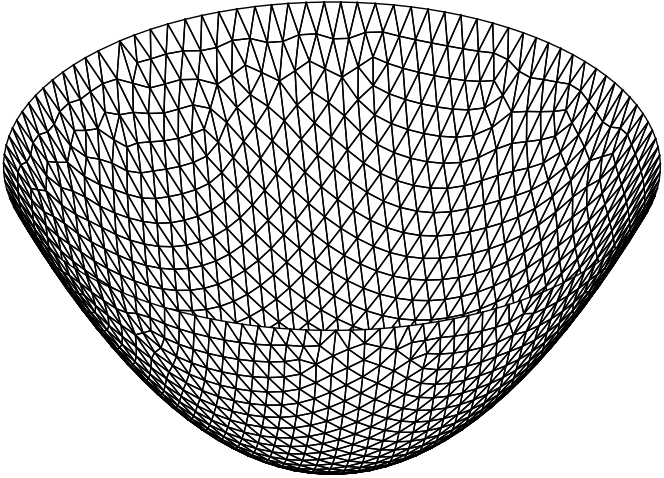}}
\caption{The triangluar mesh in the unit disk (a) and on
the parabolic surface (b).}
\label{fig:mesh-Parabolic}
\end{figure}

Finally, for the unclosed parabolic surface, we firstly generate a good quality
triangluar mesh on the unit disk, then lift the mesh nodes onto the parabolic
surface, as shown in
Fig.\,\ref{fig:mesh-Parabolic}. 

\section{SCFT on general curved surfaces}
\label{sec:scft}

In this section, we will introduce a standard field theory model for an
incompressible $AB$ diblock copolymer melt on general curved surfaces. 
We consider $n$ $AB$ diblock copolymers with the polymerization of $N$
confined on surface $\calS$ with a total surface area $|\calS|$.
The volume fraction of block $A$ is $f$ and that of block $B$ is $1-f$.
The two blocks disfavor with each other characterized by the
Flory-Huggins parameter $\chi N$. 
We assume that the statistical segment length and volume
of two blocks are equal, \textit{i.e.}, $b_A=b_B=b$ and $v_A=v_B=v_0$.
A characteristic length of the copolymer chain can be defined
by the unperturbed radius of gyration, which is used as the unit
of length so that all spatial lengths are presented in units of
$R_g=b\sqrt{N/6}$. With the incompressible melt assumption, the
average segment density is uniform in space and given by
$v_0=1/\rho_0=V/(n N)$, $\rho_0$ is the number density. Here we use a
continuous Gaussian chain to model the block
copolymer\,\cite{fredrickson2006equilibrium,matsen2002standard}.
Based on the statistical mechanics, and using
Hubbard-Stratonovich transformation and infinite dimensional
Fourier transformation\,\cite{
fredrickson2006equilibrium, jiang2015analytic,
fredrickson2002field}, one can derive the
field-based model to describe the phase behavior of the block
copolymer system. However, directly solving the field-based model
is nontrivial due to the very complicated nature of the
functional integral exhibited in the partition
function\,\cite{fredrickson2006equilibrium}. 
In order to simplify the field model,
we often use an analytic
approximation technique called the mean-field (saddle-point)
approximation which ignores the field fluctuations and assumes that
the functional integral is dominated by a single field configuration.
This approximation method become accurate when the dimensionless
chain concentration $C=\rho_0 R_g^3/N$ goes to infinity. 
In the case of high molecular weight block copolymer melts, where
$C$ can be very large, the approximation will make sense. From
the viewpoint of asymptotic theory, the functional integral,
$\textit{i.e.,}$ the partition function, can be approximated by
the integrand function at the critical point (actually the saddle
point) when the parameter $C$ becomes very large.
Then one can obtain the mean-field
model. Specifically, within the mean-field approximation,
the effective Hamiltonian of this system is
\begin{align}
  H[w_+, w_-] = &
  \frac{1}{|\calS|}\int \mathrm d\bx\, \left\{ 
  -w_+(\bx) + \frac{w_-^2(\bx)}{\chi N} 
  \right\} -\log Q[w_+(\bx), w_-(\bx)].
  \label{eq:hamiltonian}
\end{align}
In the above equation, 
$w_+(\bx)$ and $w_-(\bx)$ are the fluctuating pressure and exchange
chemical potential fields, respectively. The pressure field
enforces the local incompressibility, while the exchange chemical
potential is conjugate to the density operators.
$Q$ is the single chain partition 
functional subjected to external fields $w_+$ and $w_-$.
$\mathrm d\bx$ is the area element of the surface.
First-order variations of the free energy functional with respect
to the fields lead to the following SCFT equations,
\begin{align}
	& \frac{\delta H}{\delta w_+(\bx)} = \phi_A(\bx) + \phi_B(\bx) - 1 = 0,
	\label{eq:scftwplus}
	\\
	& \frac{\delta H}{\delta w_-(\bx)} = 
   \frac{2 w_-(\bx)}{\chi N} - [\phi_A(\bx) - \phi_B(\bx)] = 0,
	\label{eq:scftwminus}
\end{align}
where $\phi_A(\bx)$ and $\phi_B(\bx)$ are the monomer densities of block $A$
and $B$, respectively.  Then using the continuous Gaussian chain model,
\begin{align}
  & Q = \frac{1}{|\calS|}\int \mathrm d\bx\,q(\bx,1) = \frac{1}{|\calS|}\int
    \mathrm d\bx\,q(\bx,s)q^\dag(\bx,s), ~~~ \forall s \in [0,1],
  \\
  & \phi_A(\bx) =
  \frac{1}{Q}\int_0^f \mathrm ds\,q(\bx,s)q^{\dag}(\bx,s),
  \label{eq:scft:phiA}
  \\
  & \phi_B(\bx) =
  \frac{1}{Q}\int_f^1 \mathrm ds\,q(\bx,s)q^{\dag}(\bx,s).
  \label{eq:scft:phiB}
\end{align}
The forward propagator, $q(\bx,s)$,  represents
the probability weight that the chain of contour length $s$ has its end at
surface position $\bx$, where the variable $s$ is used to parameterize each
copolymer chain. $s = 0$ represents the tail of block $A$ and $s = f$ is the
junction between block $A$ and $B$.  From the continuous Gaussian chain model,
$q(\bx,s)$ satisfies the PDE
\begin{equation}
  \label{eq:MDE}
	\begin{aligned}
  \frac{\partial }{\partial s}q(\bx,s) &=
	[ \Delta_{\calS} - w(\bx,s)] q(\bx,s), 
	\\
  q(\bx,0) &= 1,
  \\
	w(\bx,s) &= \left\{   
	\begin{array}{rl}
		w_+(\bx)-w_-(\bx), &  0\leq s \leq f,
		\\
		w_+(\bx) +w_-(\bx), &  f\leq s \leq 1.
	\end{array}
\right.
	\end{aligned}
\end{equation}
The backward propagator, $q^{\dag}(\bx,s)$, satisfies an almost
identical PDE of Eqn.\,\eqref{eq:MDE}, except that it represents the
probability weight from $s=1$ to $s=0$.
In particular, it satisfies
\begin{equation}
  \label{eq:MDEplus}
	\begin{aligned}
  \frac{\partial }{\partial s}q^\dag(\bx,s) &=
	-[\Delta_{\calS} - w^\dag(\bx,s)] q(\bx,s), 
	\\
  q^\dag(\bx,1) &= 1,
  \\
	w^\dag(\bx,s) &= \left\{   
	\begin{array}{rl}
		w_+(\bx)+w_-(\bx), &  f\leq s \leq 1,
		\\
		w_+(\bx) -w_-(\bx), &  0\leq s \leq f.
	\end{array}
\right.
	\end{aligned}
\end{equation}

\section{Methodology}
\label{sec:method}

%

The equilibrium point, as well as the saddle point, of the SCFT
model corresponds to the ordered phase. And 
finding the saddle points requires iteration methods.
The previous study has been shown that the effective Hamiltonian
\eqref{eq:hamiltonian} can reach its
local minima along the exchange chemical field $w_-$ and maxima along the
pressure field $w_+$\,\cite{jiang2015analytic, fredrickson2002field}. Although
it still lacks a rigorous theoretical guarantee, the numerical behavior has
demonstrated its effectiveness. Based on this observation, we can use an
alternating direction scheme to obtain the saddle points.  Specifically, we
introduce a fictitious time variable $t$, and at each time step, the
saddle-point search method is given as
\begin{align}
	\frac{\partial }{\partial t}w_+(\bx,t) &= \frac{\delta
	H[w_+, w_-]}{\delta w_+(\bx, t)},
	\label{eq:eulerwplus}
	\\
	\frac{\partial }{\partial t}w_-(\bx,t) &= -\frac{\delta
	H[w_+, w_-]}{\delta w_-(\bx, t)}.
	\label{eq:eulerwminus}
\end{align}
Clearly, Eqns.\,\eqref{eq:scftwplus} and \eqref{eq:scftwminus} are
satisfied when Eqns.\,\eqref{eq:eulerwplus} and
\eqref{eq:eulerwminus} are stationary.

Within the standard framework of SCFT,
\textcolor{blue}{\textit{finding saddle point}} (FSP) of $w_\pm$
can be obtained by the following iteration scheme:
\begin{description}
	\item[Step 1] Given initial estimations of fields $w_\pm(\bx, 0)$.
    \item[Step 2] Compute forward and backward propagator operators
		$q(\bx,s)$ and $q^\dag(\bx,s)$ on a general curved
		surface (see Sec.\,\ref{subsec:mde}).
	\item[Step 3] Obtain $Q$, $\phi_A(\bx)$ and $\phi_B(\bx)$ by
		integral equations (see Sec.\,\ref{subsec:sintegral} and
		Sec\,\ref{subsec:surfaceintegral}), and evaluate the
		value of effective Hamiltonian $H$.
	\item[Step 4] Update fields $w_+(\bx,t)$ and $w_-(\bx,t)$ by 
		Eqns.\,\eqref{eq:eulerwplus} and \eqref{eq:eulerwminus}
		using iteration methods (see Sec.\,\ref{subsec:euler}). 
	\item[Step 5] Repeat steps 2-4 until a convergence criterion
		is met.
\end{description}

In the polymeric systems, each self-assembled microphase possesses its
characteristic scale due to the statistical length of the polymer chain.  In
the flat space, the size and shape of the computational domain can affect the
energy value of self-assembled structures\,\cite{matsen1994stable,
jiang2013discovery} and the properties of the polymer materials, such as the
mechanical property\,\cite{barrat2005introducing, tyler2003stress}.  As in the
flat domain cases, the ordered pattern formed on the curved surface has its
characteristic scale. In contrast, the surface size also affects the
self-assembled patterns.  For example, the radius of sphere affects the
self-assembled phases of diblock copolymers and their energy
values\,\cite{chantawansri2007}.  However, it still lacks a systematical method
to select an optimal size for a self-assembled structure on a general curved
surface. Here, we will propose a theoretic tool to efficiently handle this
problem, \textit{i.e.}, an adaptive approach of optimizing the size of a
general curved surface for a given ordered structure.

In order to obtain the optimal surface size, the
effective Hamiltonian of SCFT shall be viewed as a function of
the surface size, as well as a functional of the
field functions. 
Therefore, the completed optimization problem of solving SCFT
equations becomes 
\begin{align}
	\min_{\calS}\max_{w_+}\min_{w_-} H[w_+(\bx),
	w_-(\bx), \calS]. 
	\label{eq:scftOptorig}
\end{align}
\textcolor{blue}{
In practice, we parameterize the surface $\mathcal{S}$ by 
\begin{align}
	\mathcal{S}_\Gamma=\{\Gamma\cdot\bx: \bx\in\mathcal{S}_0 \},
	\label{}
\end{align}
where $\calS_0$ is the initial surface, $\Gamma>0$ is a scale factor which
describes the surface size.  For example, the parameter $\Gamma$ of a sphere is
its radius.  Then we can optimize the surface size by adjusting the scale factor
$\Gamma$.
}
Therefore the optimization problem \eqref{eq:scftOptorig} becomes
\begin{align}
	\min_{\Gamma}\max_{w_+}\min_{w_-} H[w_+(\bx),
    w_-(\bx), \Gamma]. 
	\label{eq:scftOpt}
\end{align}
It can be solved by the following iteration procedure, here we
name SCFT iteration, including finding the saddle point of SCFT
and adaptively optimizing the scale factor $\Gamma$ of a
given surface,
\begin{description}
  \item[Step 1] Give proper parameters $\chi N$, $f$, a surface $\calS$, and reasonable initial distributions of $w_{\pm}$.
  \item[Step 2] Fix $\mathcal{S}$, finding the
	  saddle point of SCFT  through FSP procedure
	  and obtaining the effective Hamiltonian.
  \item[Step 3] Fix $w_\pm$, optimizing the size of $\mathcal{S}$
	  by the method described in Sec.\,\ref{subsec:adjustBox},
	  and evaluating the value of effective Hamiltonian.
  \item[Step 4] Repeat steps 2-3 until the effective
	  Hamiltonian discrepancy is lower than a given convergence
	  criterion.
\end{description}

\subsection{Surface finite element method for PDE}
\label{subsec:mde}

In the iteration scheme outlined above, the most time consuming
step is solving the PDEs of \eqref{eq:MDE} and \eqref{eq:MDEplus}. 
For a special surface, such as a sphere, the global basis of
spherical harmonics can be used to expand the spatial functions.
Therefore the (pseudo) spherical harmonic spectral method can be
used to solve PDEs on sphere.
For a general curved surface, however, the global basis might not
exist. Therefore, the methods based on local basic functions shall be
developed to solve  SCFT model on general curved surface, such as the finite
volume or element method. In this work, we focus our attention on the
development of the finite element method. We discretize the
Laplace-Beltrami operator in \eqref{eq:MDE} and
\eqref{eq:MDEplus} by the linear surface finite element method.

Here we just present the finite element discretization for
Eqn.\,\eqref{eq:MDE}. For Eqn.\,\eqref{eq:MDEplus} of
backward propagator $q^{\dag}$, one can obtain the discretization
scheme similarly. Firstly, we rewrite \eqref{eq:MDE} into the
variational formulation: find $q$ in $H^1(\calS)$, which satisfies
\begin{equation}\label{eq:FEM}
    \left(\frac{\partial}{\partial s} q, v\right)_\calS = -(\nabla_\calS q,
    \nabla_\calS v)_\calS - (wq, v)_\calS, ~~\text{ for all } v\in H^1(\calS),
\end{equation}
where $(\cdot,\cdot)_S$ represents the $L^2$ inner product on
$\calS$. 

Replacing the infinite dimensional space $H^1(\calS)$ by the
finite dimension space $\mathcal V_h$, we obtain the linear
finite element discretization: find $q_h=\sum_{i=1}^N
q_i(s)\varphi_i(\bx)$ in $\mathcal V_h$, which satisfies
\begin{equation}\label{eq:FEM1}
    \left(\frac{\partial}{\partial s} q_h, v_h\right)_{\calS_h} =
    -(\nabla_{\calS_h}
    q_h, \nabla_{\calS_h} v_h)_{\calS_h} - (w_hq_h, v_h)_{\calS_h},
    ~~\text{ for all } v_h\in \mathcal V_h, 
\end{equation}
where $(\cdot, \cdot)_{\calS_h}$ represents the $L^2$ inner product on the
triangular mesh $\calS_h$, and $w_h(\bx, s)=\sum_{i=1}^N
w(\bx_i, s)\varphi_i(\bx)$ is the linear interpolation of $w(\bx, s)$.
Replacing $v_h$ by $\varphi_j$, $j=1, 2, \cdots, N$, we can have the matrix
form of
\eqref{eq:FEM1}, 
\begin{equation}
    M\frac{\partial}{\partial s}\mathbf q(s) = - (A + F) \mathbf q(s), 
\end{equation}
where 
\begin{equation*}
    \mathbf q(s) = \left(q_1(s), q_2(s), \cdots, q_N(s)\right)^t, 
\end{equation*} 
and 
\begin{equation*}
    M_{i,j} = (\varphi_i, \varphi_j), A_{i, j} = (\nabla_\calS\varphi_i,
    \nabla_\calS\varphi_j), F_{i, j} = (w_h\varphi_i, \varphi_j).
\end{equation*}
For the discretization of $s$-direction, we use the Crank-Nicolson method,
\begin{equation}\label{eq:FEMs}
    M \frac{\mathbf q^{n+1} - \mathbf q^n}{\Delta s} = - \frac{1}{2}(A + F)\left[\mathbf
    q^{n+1} + \mathbf q^{n}\right],
\end{equation}
where $\Delta s$ is the time step size. It should be noted that any
stable time discretization scheme can be used here.
Combining with the above two steps, we obtain the full
discretization scheme
\begin{equation}\label{eq:FEMs}
    \left[M + \frac{\Delta s}{2} (A+F)\right] \mathbf q^{n+1} = 
	\left[M - \frac{\Delta s}{2}(A + F)\right]\mathbf{q}^{n} .
\end{equation}
If the linear algebraic system \eqref{eq:FEMs} is relatively
small, we solve it with the direct method. Otherwise the algebraic
multigrid method can be applied\,\cite{chenlong}. \textcolor{blue}{Furthermore, the order of the
above discretizations for the approximation to the solution of the modified
diffusion equations is $O(\Delta s^2 + h^2)$, where $h$ is
triangular surface mesh size\,\cite{Dziuk2007}}.

\subsection{Adaptively optimizing the curved surface}
\label{subsec:adjustBox}

When fixed fields $w_\pm(\bx)=w_\pm^f(\bx)$, the optimization
problem of \eqref{eq:scftOpt} becomes 
\begin{align}
    \min_{\Gamma}H[w_+^f(\bx), w_-^f(\bx), \Gamma].
	\label{eq:scftoptival}
\end{align}
Any appropriate optimization method can be chosen to solve this
problem, such as the simplest steepest descend method,
\textit{i.e.},
$d\Gamma/dt = -dH/d\Gamma$.
However, due to the complexity of SCFT, it is hardly written
down the derivative of the
SCFT's Hamiltonian with respect to the scale factor of a 
curved surface $d H/d \Gamma$ analytically.
Alternatively, we can approximate $d H/d \Gamma$ numerically
through the finite difference method. 
To improve the effectiveness of algorithm, in practical
implementation, we adapt the nonlinear conjugate-gradient
(CG) method to minimize objective
function\,\eqref{eq:scftoptival}\,\cite{nocedal2006numerical},
\textit{i.e.},
\begin{align}
	\Gamma^{k+1} = \Gamma^{k} + \alpha^k d^k,
  \label{eq:adjustR}
\end{align}
where $d^k$ is the conjugate gradient direction in $k$th step.
$\alpha_k$ is the step size obtained by the linear search
approach. Meanwhile, the restart technology has been used in
nonlinear CG method to avoid the oscillation and improve the
convergent rate.

\subsection{Iteration method of finding saddle point}
\label{subsec:euler}

The iteration methods to update the fields 
are dependent on the mathematical structure of SCFT. 
An important fact is that the effective Hamiltonian
\eqref{eq:hamiltonian} of diblock copolymers can reach its local
minima along the exchange chemical field $w_-(\bx)$, and reach
the maxima along the pressure field
$w_+(\bx)$\,\cite{fredrickson2006equilibrium}. 
\textcolor{blue}{
Then a series of accelerate convergence methods, such as the
semi-implicit method, can be designed to find the saddle point.
More general results for multicomponent polymeric systems with
more than two chemically distinct blocks, including a simple
analysis of SCFT model and a class of efficient iteration methods
of finding saddle points, can be found in
Ref.\,\cite{jiang2015analytic}.  
It should be noted that the semi-implicit scheme is
obtained through the asymptotic expansion and Fourier transformation. 
Based on similar techniques, it can be extended to sphere
surface problems using spherical harmonic
transformation\,\cite{chantawansri2007}.
However, it is hardly to find a representation of spatial
functions defined on a general curved surface using global basis.
}
Therefore we choose the alternative direction explicit Euler method to
update the fields $w_\pm(\bx)$. Specifically, this
approach is expressed as 
\begin{equation}
	\begin{aligned}
		w_+^{k+1}(\bx) &= w_+^k(\bx) + \lambda_+
		[\phi_A^k(\bx)+\phi_B^k(\bx)-1 ],
		\\
		w_-^{k+1}(\bx) &= w_-^k(\bx) - \lambda_-
   \left(\frac{2 w_-^k(\bx)}{\chi N} - [\phi_A(\bx) - \phi_B(\bx)]
   \right).
	\end{aligned}
	\label{}
\end{equation}

\subsection{Integral formula along $s$-direction }
\label{subsec:sintegral}

A modified numerical integration formula for a closed interval is
chosen to evaluate integral equations
\eqref{eq:scft:phiA}-\eqref{eq:scft:phiB} that can guarantee
fourth-order precision in $s$-direction whether the number of
discretization points is even or odd\,\cite{press1992numerical}.
\begin{align}
	\int_{0}^{s_n} \mathrm ds\,f(s) = \Delta s \left\{
	-\frac{5}{8}(f_0+f_{n}) 
	+\frac{1}{6}(f_1+f_{n -1}) 
	-\frac{1}{24}(f_2+f_{n-2}) 
	+\sum_{j=0}^{n} f_{j} 
\right\} + O(\Delta s^4),
\end{align}
where $n$ is the number of discretization nodes along
$s$-direction, $\Delta s$ is a constant step, and $s_n=n\Delta s$. 
$f_k = f(k\Delta s)$, $k=0,1,\dots, n$.

\subsection{Surface integral}
\label{subsec:surfaceintegral}

Given a linear finite element function $v_h = \sum_{i=1}^N
v_i\varphi_i(\bx)$ on $\calS_h$, we can calculate the integration of
$v_h$ on $\calS_h$ from Eqn.\,\eqref{eq:varphi}, 
\begin{equation*}
    \int_{\calS_h} v_h \, \mathrm d \bx = \sum_{\tau_h\in\mathcal T_h} \int_{\tau_h}
    v_i\varphi_i(\bx) + v_j\varphi_j(\bx) + v_k\varphi_k(\bx)\, \mathrm d \bx = \sum_{\tau_h\in\mathcal T_h} \frac{v_i + v_j +
    v_k}{3}|\tau_h|, 
\end{equation*}
where $\varphi_i(\bx)$ is a linear function about $\bx$ on each
triangle face, $v_i$, $v_j$ and $v_k$ are the function
values on the three vertices of a triangle $\tau_h = (\bx_i,
\bx_j, \bx_k)$, respectively, and $|\tau_h|$ is the area of $\tau_h$.
\textcolor{blue}{ Notice that, for linear function on plain
triangle, the above single point numerical integration formula is
accurate. Given a function $v\in H^1(\calS)$ and its interpolation
$v_I$ on $\calS_h$, the error estimate can be obtained by
Cauchy-Schwarz inequality and interpolation error estimate\,\cite{Dziuk1988}: 
\begin{equation*}
    \left|\int_{\calS_h}\bar v \, \mathrm d\bx - \int_{\calS_h} v_I\, \mathrm d
    \bx\right| \leq \int_{\calS_h} \left|\bar v - v_I\right|\, \mathrm d\bx 
    \leq |\calS_h|^{\frac{1}{2}} ||\bar v - v_I||_{0, \calS_h}
    \lesssim h^2 ||v||_{1, \calS},
\end{equation*}
where $\bar v(\bx) = v(\mathcal P(\bx))$ and $|\calS_h|$ is the area of
$\calS_h$. 
}

\section{Numerical Results}
\label{sec:rslt}

In this section, we will use five closed curved surfaces
including sphere, torus, double torus, heart, orthocircle, and an
unclosed parabolic surface to demonstrate the efficiency of our
proposed numerical method. To ensure the accuracy, we set the substep of
contour length $s$ to $0.005$ in the following simulations. The
convergence criterion of FSP is 
\begin{align}
	\max\left\{
	\left\|\frac{\delta H}{\delta w_+} \right\|_{\ell^\infty},
	\left\|\frac{\delta H}{\delta w_-} \right\|_{\ell^\infty} 
	\right\} \leq 10^{-6},
	\label{}
\end{align}
and that of SCFT iteration is the change of effective Hamiltonian
smaller than $10^{-6}$.

\subsection{Efficiency of the numerical method}

\begin{figure}[H]
\setlength{\captionmargin}{2pt}
\centering
    \subfigure[]{\includegraphics[scale=0.55]{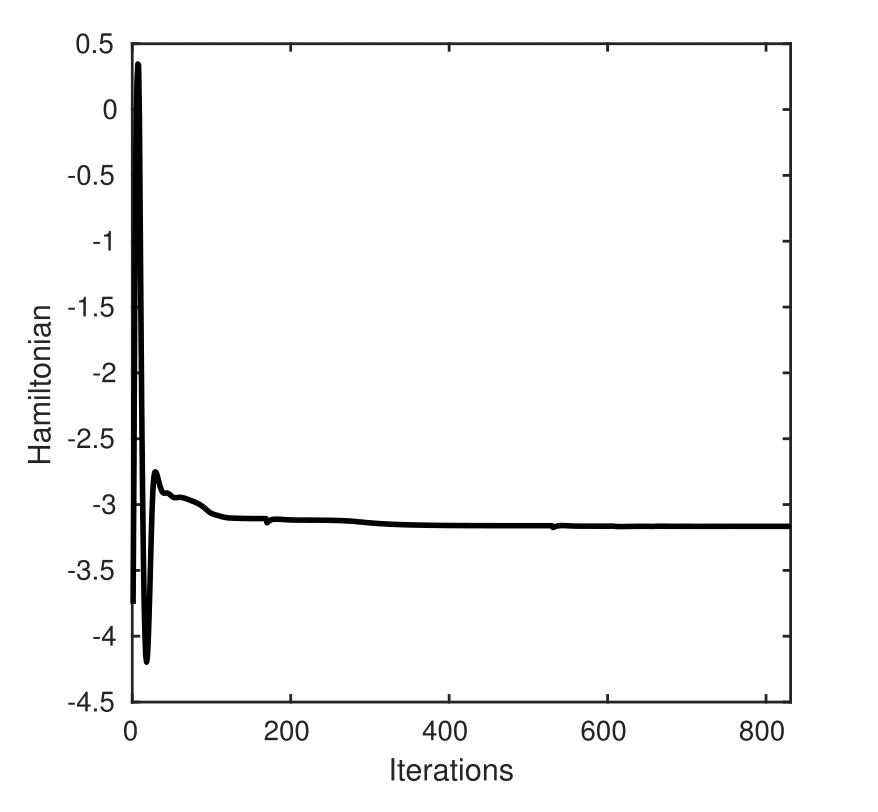}}
    \hspace{0.0cm}
\subfigure[]{\includegraphics[scale=0.55]{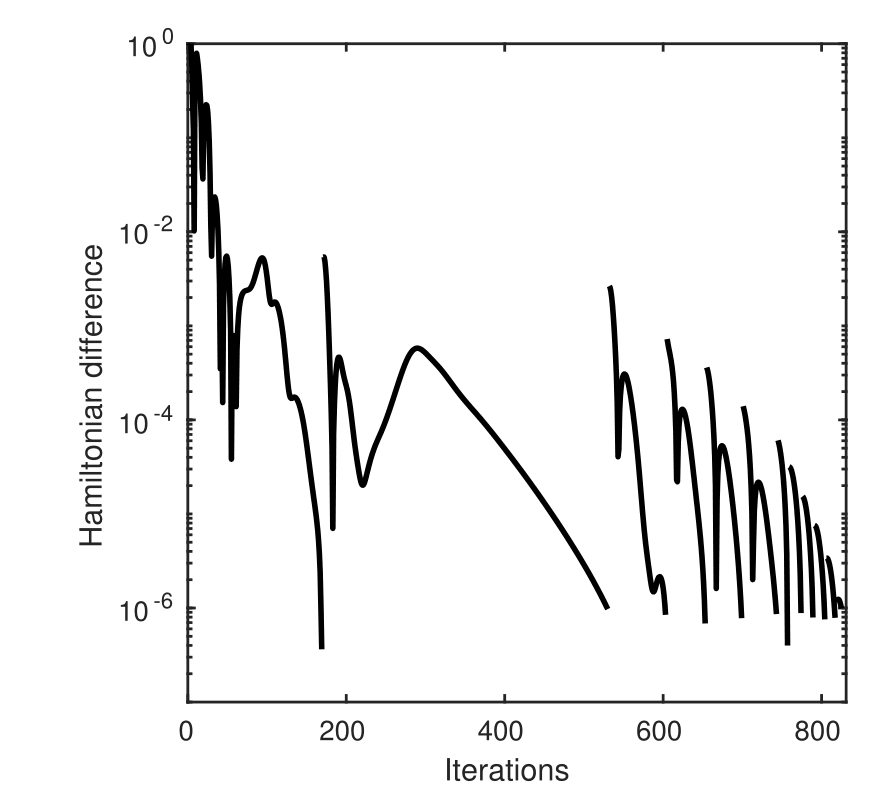}}
\    
\caption{
       (a) the values of effective Hamiltonian and (b) the
	   discrepancy of Hamiltonian respective to iteration of the
	   simulation when computing 12-microdomain phase on the sphere
	   with 10242 nodes and 20480 elements.
		}
		\label{fig:sphere-5}
\end{figure}

Since it is hard to obtain a nontrivial analytical solution of
SCFT on a general curved surface, we have to verify the
convergence of the numerical solution of our proposed method by
running the simulations on a series of subdivision meshes. Here
we take the 12-microdomain spots structure on the sphere with
model parameters $\chi N=30.0$, $f=0.20$, $\Gamma=2.9$ as an
example to show the convergence of our method, see
Tab.\,\ref{tab:refinemesh}. 
\begin{table}[H]
\caption{
The calculated information of SCFT simulations on a series of subdivision
meshes, where ``--'' means divergence.}
  \label{tab:refinemesh}
  \centering
\begin{tabular}{|c|c|c|c|c|c|}
 \hline
 \# of Node  & \# of Elem. & $\Gamma$ & $\Gamma$-diff. & Hamiltonian &
 Ham. diff. \\
 \hline
162  & 320   & -- & -- & -- &    \\ 
 \hline
 642   & 1280  & 3.707  & 8.07e-1 & -3.173  &  \\
 \hline
2562  & 5120  & 3.579  & 1.29e-1 & -3.161 &  1.22e-2\\
 \hline
10242 & 20480 & 3.566  & 1.13e-2 & -3.165& 4.89e-3\\
 \hline
40962 & 81920 & 3.563  & 2.40e-3 & -3.167 & 1.42e-3 \\
 \hline
\end{tabular}
\\
\end{table}
In these series of simulations, the initial fields on each node
$(x, y, z)$ is $ w_{-} = \chi N\sin(5\theta)$, $w_{+} = 0 $,
where $\theta = \arctan(y/x) \in [0, 2\pi]$.  The initial surface
is a unit sphere with scale factor $\Gamma=2.9$.  We uniformly
refine the initial mesh and the number of nodes and triangular
elements can be found in the first and second columns of
Tab.\,\ref{tab:refinemesh}. From the table, one can find that,
along with the refinement of the mesh, the value of effective
Hamiltonian is indeed convergent, and furthermore, the
scale factor $\Gamma$ is adaptively optimized and convergent. The
finally convergent morphology can be found in
Fig\,\ref{fig:sphere} (a). In order to further demonstrate the
iteration procedure, we give the iteration details of the SCFT
iteration when using $10242$ nodes and $20480$ elements as shown
in the fifth row of Tab.\,\ref{tab:refinemesh}.  
The information is given in Tab.\,\ref{tab:fixedmesh}.  
The effective Hamiltonian and its discrepancy with respect to
iteration are given in Fig.\,\ref{fig:sphere-5}. It costs $15$
SCFT iterations, and
$831$ FSP iterations. The scale factor $\Gamma$ is optimized
from $2.9$ to $3.5656$.  Correspondingly, the discrepancy of the
effective Hamiltonian values is $5.9\times 10^{-2}$ (from
$-3.106$ to $-3.165$). The change amount of effective Hamiltonian
is enough to determine the thermodynamic stability, as well as
the phase boundary in the phase diagram, of self-assembled
structures in block copolymers\,\cite{chantawansri2007,
fredrickson2006equilibrium, matsen1994stable,
jiang2013discovery}.  The results are consistent with the
previous theoretical results\,\cite{chantawansri2007}.  However,
it should be noted that, in \cite{chantawansri2007}, the
authors obtained the accurate effective Hamiltonian values
through manually changing the scale factor $\Gamma$ for each
time. Here, $\Gamma$ is optimized adaptively.

\begin{table}[H]
  \caption{
  The iteration details of the SCFT simulation when computing the
  12-microdomain spotted phase on the sphere surface used 
  10242 nodes and 20480 elements.
  }
  \label{tab:fixedmesh}
  \centering
  \begin{tabular}{|c|c|c|c|c|c|}
    \hline
	Iter. of SCFT & Iters. of FSP & $\Gamma$  &
	$\Gamma$-Diff. & Hamiltonian & Ham. Diff.  \\
    \hline
	1 & 169 & 2.90 & - & -3.10648180 & -
	\\ \hline
	2 & 361 & 3.372922 & 4.729e-1 & -3.16079157  & 5.431e-2
	\\ \hline
	3 & 73 & 3.656954 & 2.843e-1  & -3.16437826  & 3.587e-3
	\\ \hline
	4 & 50 & 3.530368 & 1.266e-1  & -3.16530810  & 9.298e-4
	\\ \hline
	5 & 46 & 3.580692 & 5.032e-2  & -3.16538583  & 7.773e-5
	\\ \hline
	6 & 44 & 3.559392 & 2.130e-2  & -3.16543307  & 4.723e-5
	\\ \hline
	7 & 14 & 3.568215 & 8.823e-3  & -3.16529309  & 1.399e-4
	\\ \hline
	8 & 17 & 3.564173 & 4.043e-3  & -3.16546705  & 1.739e-4 
	\\ \hline
	9 & 15 & 3.566281 & 2.018e-3  & -3.16540431  & 6.273e-5
	\\ \hline
    10 & 15 & 3.565232 &1.049e-3  & -3.16543986  & 3.554e-5 
	\\ \hline
    11 & 13 & 3.565744 &5.120e-4  &  -3.16542586 & 1.399e-5
	\\ \hline
	12 & 8 & 3.565524 & 2.212e-4  & -3.16542759  & 1.737e-6 
	\\ \hline
    13 & 2 & 3.565564 & 4.027e-5  &  -3.16543001 & 2.422e-6 
	\\ \hline
    14 & 2 & 3.565581 & 1.061e-6  & -3.16543102  & 1.008e-6
    \\\hline
    15 & 2 & 3.565591 & 1.000e-6  & -3.16543136  & 3.377e-7
    \\ \hline    
  \end{tabular}
\end{table}

\subsection{Patterns on several curved surfaces}
\label{subsec:pattern}

In this subsection, we apply our numerical method to several
different curved surfaces, including sphere, torus, double torus,
heart, orthocircle and parabolic surfaces.  Since the SCFT is a
high nonlinear system and has multi-solutions, initial values
play an important role in determining the final morphologies and
accelerating convergent speed.  Starting from random initial
conditions usually does not obtain energetically favorable
patterns. In order to obtain these stable solutions, preferred
structures are usually chosen as initial values in the simulations.

\subsubsection{Sphere}
\label{subsubsec:sphere}

\begin{figure}[H]
\setlength{\captionmargin}{2pt}
\centering
    \subfigure[]{\includegraphics[scale=0.27]{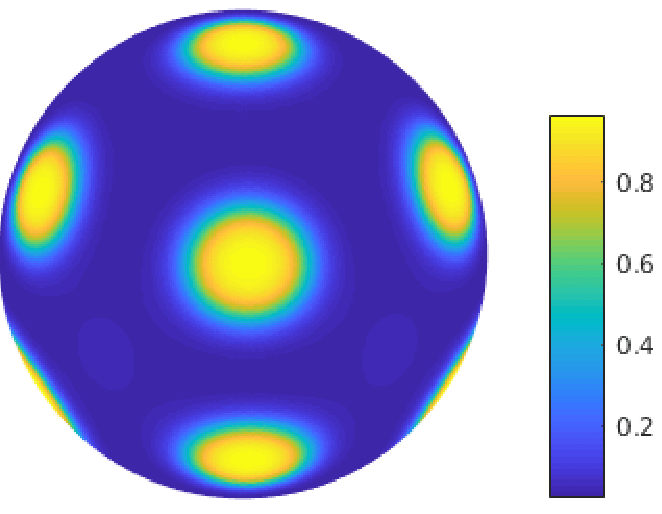}}
    \hspace{0.5cm}
\subfigure[]{\includegraphics[scale=0.30]{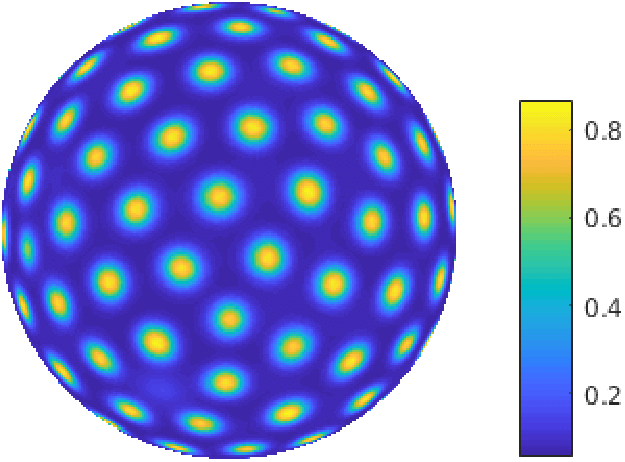}}
\\
\subfigure[]{\includegraphics[scale=0.30]{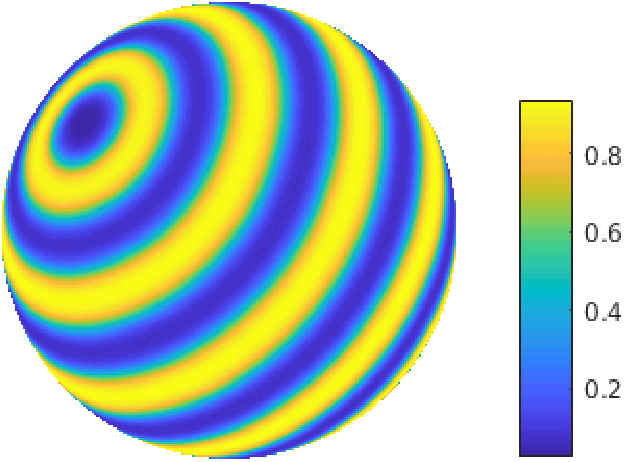}}
    \hspace{0.5cm}
\subfigure[]{\includegraphics[scale=0.30]{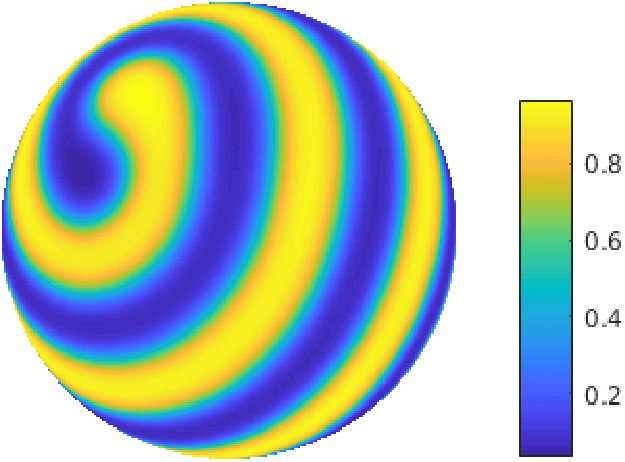}}
    \hspace{0.5cm}
\subfigure[]{\includegraphics[scale=0.30]{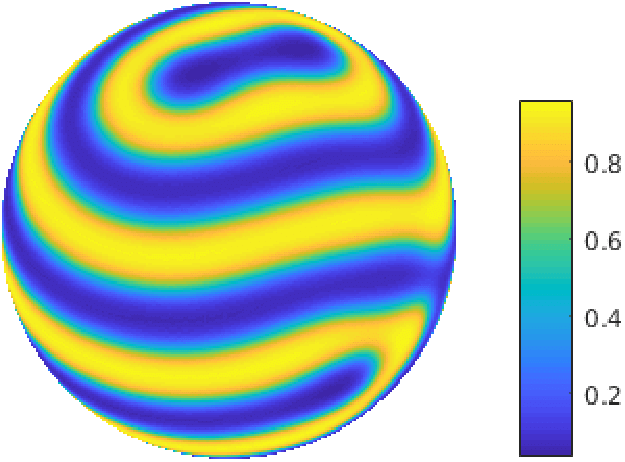}}
\caption{\label{fig:sphere}
The self-assembled patterns on the sphere obtained through
linear surface finite element SCFT simulations.  
Yellow colors correspond to large A-segment fractions.
(a). A 12-microdomain spotted phase when $\chi N=30.0$, $f=0.20$. 
(b). A 116-mircodomain spotted pattern where $\chi N=25.0$, $f=0.20$.
(c)-(e) are three striped configurations density
composition profiles when $\chi N=15.0$, $f=0.50$.
(c). A ring-form phase.
(d). A single spiral ribbon.
(e). A semi-ring striped phase.
}
\end{figure}

We firstly present the SCFT results on the sphere surface.
To capture these self-assembled structures precisely, we use
enough discretization nodes in these simulations. The final
equilibrium ordered structures are represented in
Fig.\,\ref{fig:sphere}. The initial scale factor $\Gamma_0$ and
optimal $\Gamma_{opt}$ for each self-assembled pattern are given
in Tab.\,\ref{tab:sphere:radius}.  

For the asymmetric composition of $AB$ diblock copolymers with $f=0.20$, spotted
patterns are energetically favorable.  To ensure the accuracy, $10242$ nodes
and $20480$ elements have been used to calculate these asymmetric cases.  In
contrast to the case in 2D flat space where only hexagonal pattern is globally
stable, various spotted patterns appear on the sphere surface. As shown in
Fig.\,\ref{fig:sphere} (a), a 12-microdomain spotted phase whose
spots locate at vertices of a regular icosahedron. The initial
fields of this structure have been given in the above subsection.
Then we enlarge the scale factor $\Gamma$
to $11.65$ and use random initial fields.  An ordered pattern, with 116-microdomain
spots, as shown in Fig.\,\ref{fig:sphere} (b), appears in the simulation.  It
has hexagonal lattice structure along with a small number of pentagonal
patterns.  Besides Fig.\,\ref{fig:sphere} (a) and (b), many spotted phases with
different number of microdomains have been also discovered in our
simulations. Tab.\,\ref{tab:sphere:spotNum} gives the
optimal scale factor of sphere and its corresponding number of
microdomains of a spotted phase when $\chi N=25.0$ and $f=0.20$. 
From this table, one can find that, for spotted patterns, the
number of microdomains of a spotted structure is linearly
dependent on the sphere surface area.

\begin{table}[H]
	\caption{The initial scale factor $\Gamma_0$ and optimal
	$\Gamma_{opt}$ of the sphere surface for different ordered
	pattern as shown in Fig.\,\ref{fig:sphere}.}
  \label{tab:sphere:radius}
  \centering
\begin{tabular}{|c|c|c|c|c|c|}
 \hline
 & Fig.\,\ref{fig:sphere} (a) & Fig.\,\ref{fig:sphere} (b) & Fig.\,\ref{fig:sphere} (c) & Fig.\,\ref{fig:sphere} (d) & Fig.\,\ref{fig:sphere} (e)
 \\ \hline
$\Gamma_0$ & 2.9 & 11.65 & 8 & 7.273 & 8
 \\ \hline
 $\Gamma_{opt}$ & 3.565591 & 10.529 & 7.273 & 5.468 & 6.408 
 \\ \hline
\end{tabular}
\end{table}

\begin{table}[H]
\caption{The relationship between the number of microdomains of a
spotted pattern and the optimal scale factors of sphere when $\chi N=25.0$ and
$f=0.20$. }
  \label{tab:sphere:spotNum}
  \centering
\begin{tabular}{|c|c|c|c|c|c|c|c|c|c|}
 \hline
 $\Gamma_{opt}$ & $3.430$ & $4.593$ & $5.136$ & $5.253$ & $7.411$ & $8.384$ & $9.271$ & $9.841$ & $10.53$
 \\ \hline
 \# of Spots & $12$ & $22$ & $28$ & $29$ & $56$ & $72$ & $89$ & $101$ & $116$
 \\ \hline
\end{tabular}
\end{table}

For the symmetric diblock copolymers, \textit{i.e.}, $f=0.50$, a
two-component alternating flat lamellar phase is energetically
favorable. On the sphere surface, however, various striped
patterns can be formed dependent on the initial fields and the
sphere size. As shown in Fig.\,\ref{fig:sphere} (c-e), the standard 
ring and spiral patterns are formed.
In these cases, $2560$ nodes and $5120$ elements have been used
to calculate these symmetric cases. Among these patterns,
Fig.\,\ref{fig:sphere} (c) is composed of ordinary ring-form
ribbons with point defects at only two opposite site on the
sphere surface.  The pattern in Fig.\,\ref{fig:sphere} (d) is a
single spiral ribbon crawling on the sphere surface with two
defects.  The initial fields for the classical striped phase
Fig.\,\ref{fig:sphere} (c) and a single spiral ribbon
Fig.\,\ref{fig:sphere} (d)  is $w_{-} = \chi
N\sin(8\theta+8\phi), w_{+} = 0$, where $\theta=\arctan(y/x)$,
$\phi=\arctan(z/y) \in [0, 2\pi]$, but with different initial scale factor
$\Gamma_0$ given in Tab.\,\ref{tab:sphere:radius}. It is obvious that the scale
factor dominates the morphologies of striped phases.  Fig.\,\ref{fig:sphere}
(e) is a semi-ring striped phase whose initial fields is chosen randomly.
More abundant spiral patterns on sphere surface have been found in
\cite{li2006self} with a large range of $\Gamma$. The stability of
striped patterns when $\chi N =12.5$ and $f=0.50$ has been presented by
Chantawansri \textit{et al.}\,\cite{chantawansri2007} when $\Gamma=3.1 \sim 4.9$.
However, due to the complexity of these striped patterns, more
systematical studies are still absent. 

\subsubsection{Torus}
\label{subsubsec:torus}

\begin{figure}[H]
\setlength{\captionmargin}{2pt}
\centering
\subfigure[]{\includegraphics[scale=0.26]{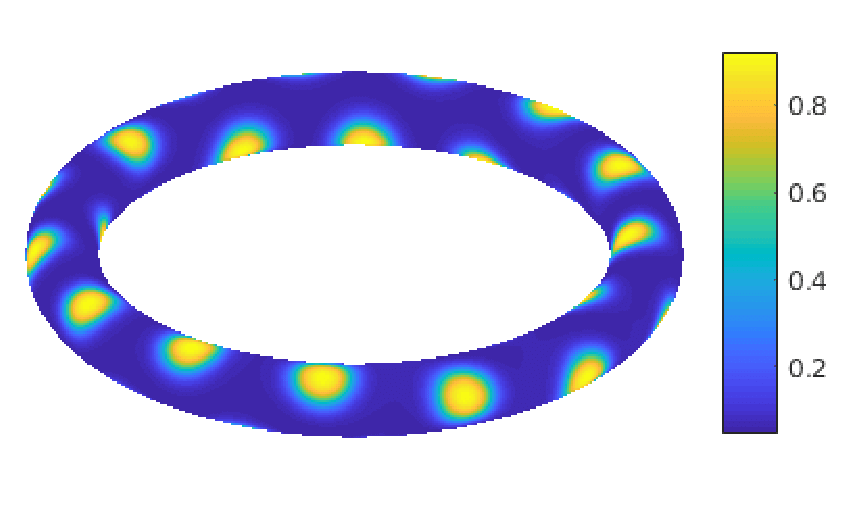}}
    \hspace{0.5cm}
\subfigure[]{\includegraphics[scale=0.26]{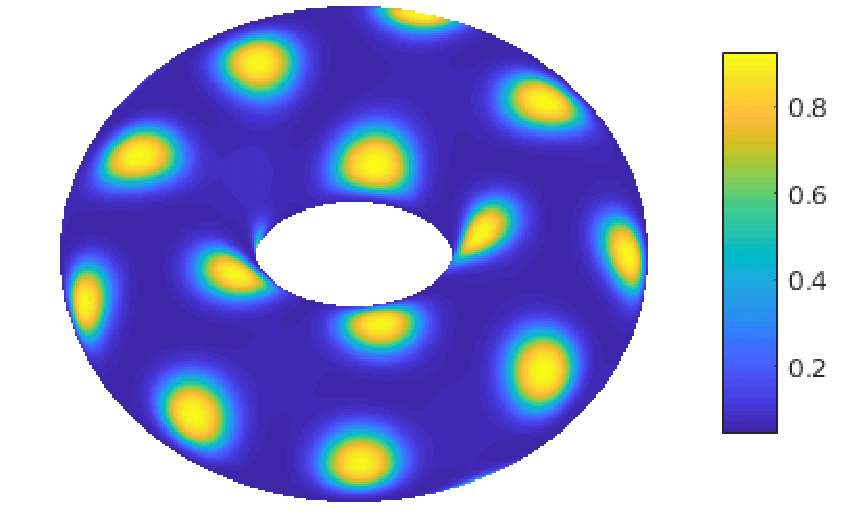}}
    \hspace{0.5cm}
\subfigure[]{\includegraphics[scale=0.26]{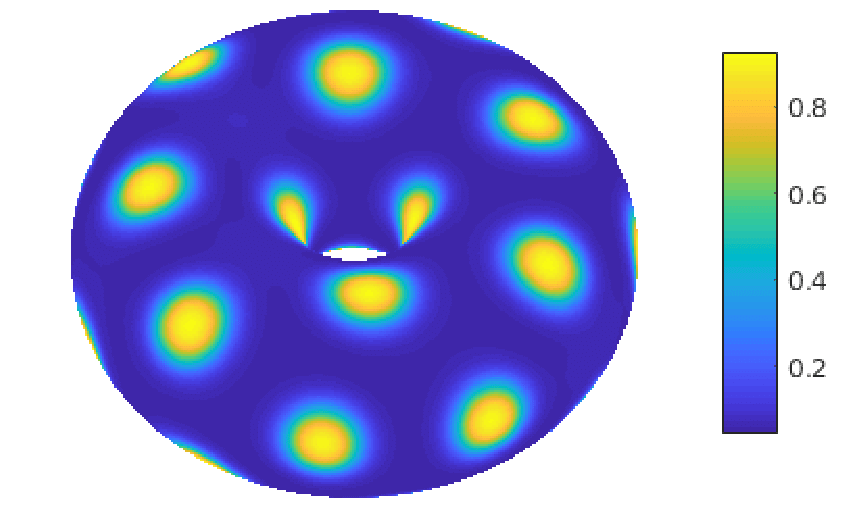}}
\\
\subfigure[]{\includegraphics[scale=0.26]{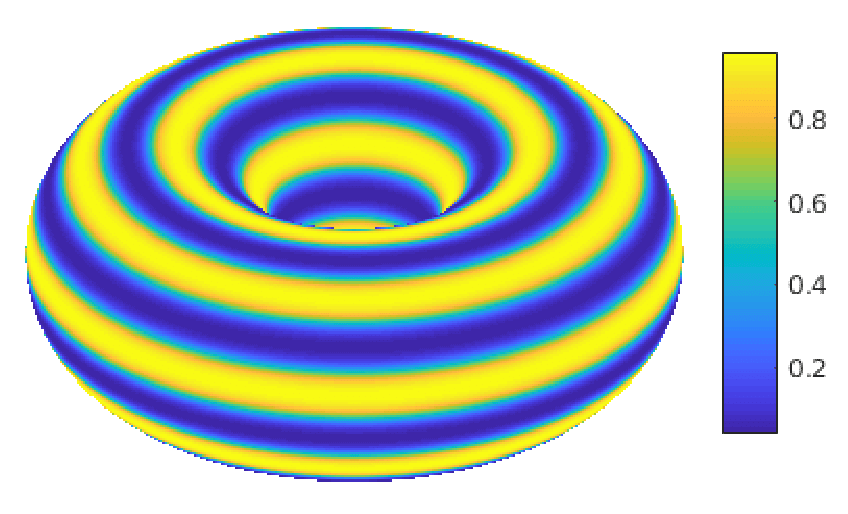}}
    \hspace{0.5cm}
\subfigure[]{\includegraphics[scale=0.26]{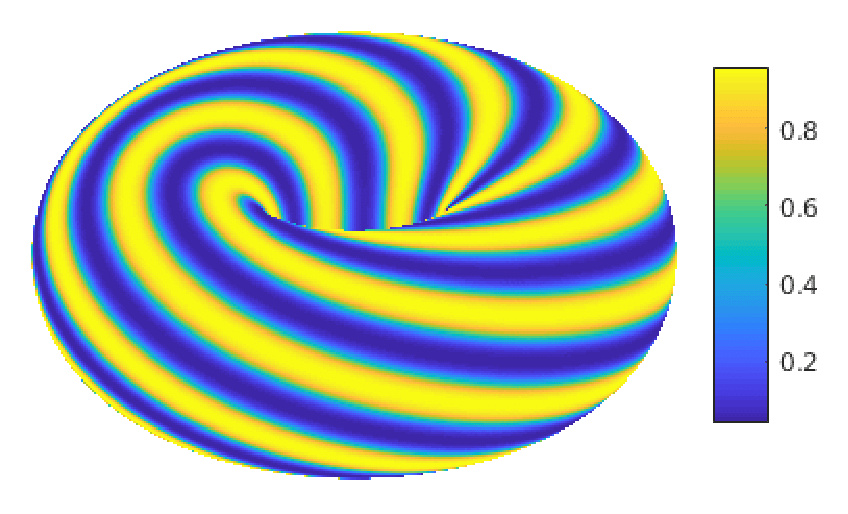}}
    \hspace{0.5cm}
\subfigure[]{\includegraphics[scale=0.26]{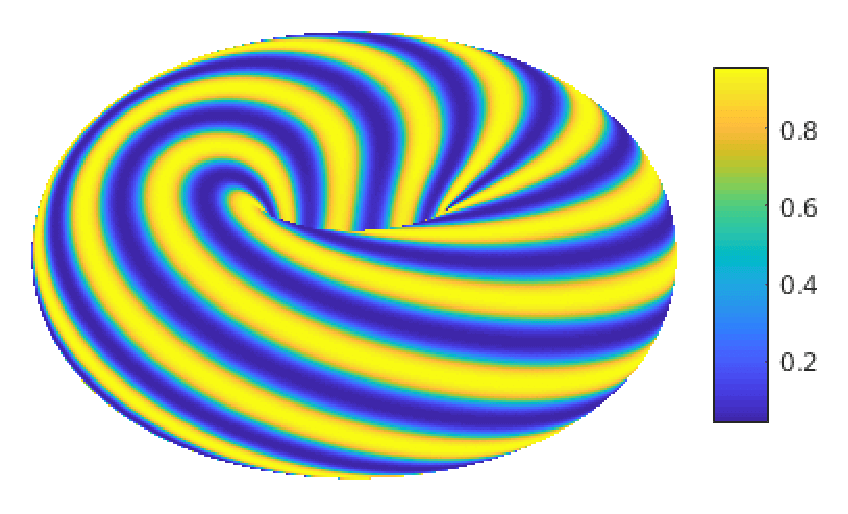}}
    \hspace{0.5cm}
\subfigure[]{\includegraphics[scale=0.26]{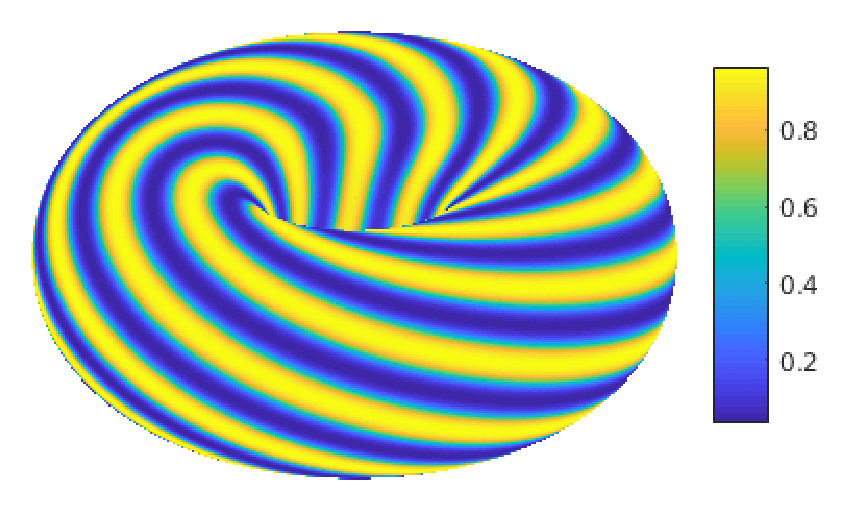}}
\caption{The self-assembled patterns on torus surface are
obtained through linear surface finite element SCFT calculations.
Yellow colors correspond to large A-segment fractions.
(a)-(c) are spotted patterns where $\chi N= 25.0$, $f=0.20$.
(d)-(g) are striped phases where $\chi N=16.0$, $f=0.50$.
The information about torus can be 
found in Tab.\,\ref{tab:radius:torus}.  }
\label{fig:torus}
\end{figure}

We next show the SCFT results on the torus surface. For every
simulation, we fix the rate of the major and minor radius of torus
surface $\gamma=R/r$ and set the initial scale factor $\Gamma_0$ as 1.  The
initial $R_0$, $r_0$ and optimal scale factor $\Gamma_{opt}$ for each ordered
pattern are presented in Tab.\,\ref{tab:radius:torus}.  The converged
self-assembled structures are given in Fig.\,\ref{fig:torus}. 

\begin{table}[H]
\caption{
    The initial $R_0$, $r_0$ and optimal scale factors $\Gamma_{opt}$ for 
different ordered structure on the torus surface. }
  \label{tab:radius:torus}
  \centering
\begin{tabular}{|c|c|c|c|c|c|c|c|}
 \hline
 & Fig.\,\ref{fig:torus} (a) & Fig.\,\ref{fig:torus} (b)
    & Fig.\,\ref{fig:torus} (c)& Fig.\,\ref{fig:torus}(d) &
    Fig.\,\ref{fig:torus}(e) &   Fig.\,\ref{fig:torus}(f) &
	Fig.\,\ref{fig:torus}(g) 
    \\ \hline
    $\gamma$ & 8 & 2 & 1.389 &2 &2 & 2 & 2
    \\ \hline
    $R_0$     & 8  & 4 & 3.333 & 4 & 4 & 4 & 4
    \\ \hline
    $r_0$     & 1  & 2 & 2.4 & 2 & 2 & 2 & 2  
	\\ \hline
	$\Gamma_{opt}$ & 1.066 & 1.044 & 1.045 & 0.969 & 1.023 & 1.119 & 1.223
    \\ \hline 
\end{tabular}
\end{table}

For asymmetric $AB$ diblock copolymers, the spotted patterns
appear, as shown in Fig\,\ref{fig:torus} (a)-(c) where $\chi N =
25.0$ and $f=0.20$.  51200 nodes and 102400 elements are used in
these calculations.
Using random initial conditions, the spotted phases with
different radii can be found as shown in Figs.\,\ref{fig:torus} (a) and
(b). To test the influence of initial condition for spotted
phases, we use two ways to choose initial condition. 
For Fig.\,\ref{fig:torus} (c), the first initial
condition is the deterministic configuration $w_{-} = 0.5(\chi
N\cos(3\theta)\cos(3\phi)+\cos(6\phi)), w_{+} = 0$, and the second is
also the random initial condition.  The same spotted patterns surprisingly
appear after calculating. It indicates that the morphologies of spotted phases
of asymmetric diblock copolymers on torus may be mainly dominated by the geometry
of the surface, \textit{i.e.}, radii $R$ and $r$ of torus, and $f$, $\chi N$.

For the symmetric diblock copolymer systems with $f=0.50$, the
equilibrium striped phases can be obtained, as shown in
Fig.\,\ref{fig:torus} (d)-(g) where $\chi N = 16.0$.  
In these calculations, $64000$ nodes and $128000$ elements are used.
From simulations, we find that striped phases are
sensitive to initial values which is different from spotted
ones. In order to obtain different striped structures,
the initial configuration is chosen as 
\begin{align*} w_-
	=\chi N\sin(n_\phi \phi + n_\theta \theta), ~~~ w_+ = 0,
\end{align*}
with different $n_\phi$ and $n_\theta$, where $\phi, \theta\in[0,2\pi]$ are the parameter coordinates of
torus surface. If we choose the initial condition with $n_\phi=0$, but with
different $n_\theta$, the unconnected stripes are obtained. For
example, when $n_\theta = 8$, the stripes of Fig.\,\ref{fig:torus} (d) emerge.
If the initial condition is chosen such that $n_\phi \neq 0$, we
can find equilibrium connected striped patterns with periods
$n_\theta = n_\phi = 6, 7, 8$, as shown in the
Fig.\,\ref{fig:torus} (e)-(g). The results are consistent with
previous calculations\,\cite{li2014mean}.

\subsubsection{Three general closed surfaces}
\label{subsubsec:generalsurface}

To further demonstrate the power of our proposed method, we next show the
results of the SCFT simulations on the double torus, heart, and orthocircle
surfaces in turn.  The parameter setting is $\chi N=25.0$, $f=0.20$ for
asymmetric diblock copolymer system, and $\chi N = 15.0$, $f=0.50$ for
symmetric case $f=0.50$.  The number of nodes (elements) used on double torus,
heart and orthocircle surfaces is $20642$ ($41288$), $172482$ ($344960$), and
$21964$ ($43952$), respectively.

\begin{figure}[H]
\setlength{\captionmargin}{2pt}
\centering
\subfigure[]{\includegraphics[scale=0.30]{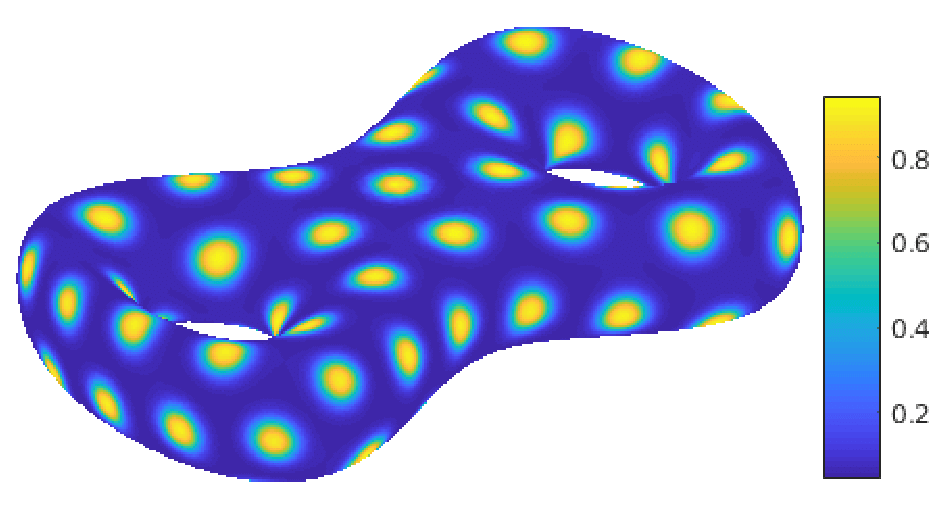}}
    \hspace{0.5cm}
\subfigure[]{\includegraphics[scale=0.30]{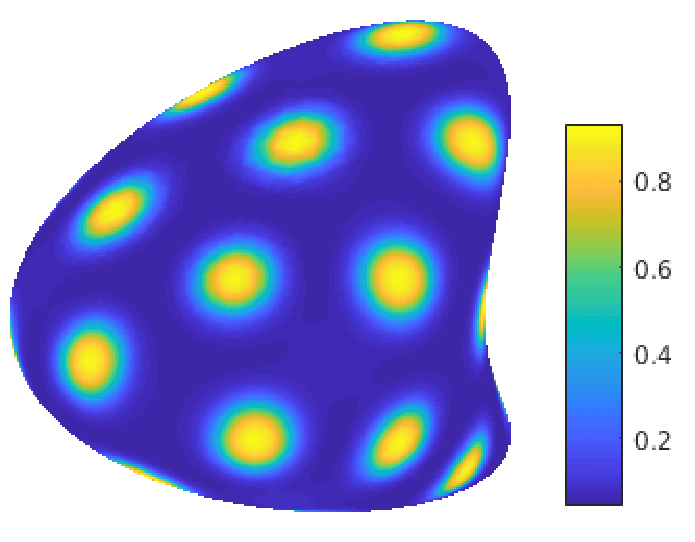}}
\hspace{0.5cm}
\subfigure[]{\includegraphics[scale=0.30]{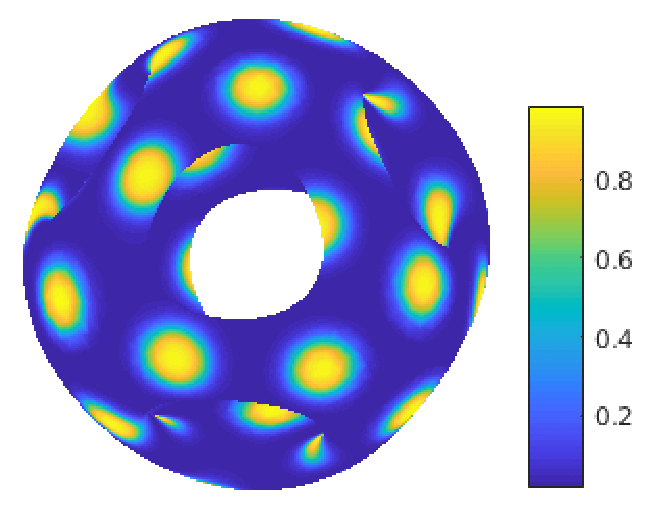}}
\\
\subfigure[]{\includegraphics[scale=0.30]{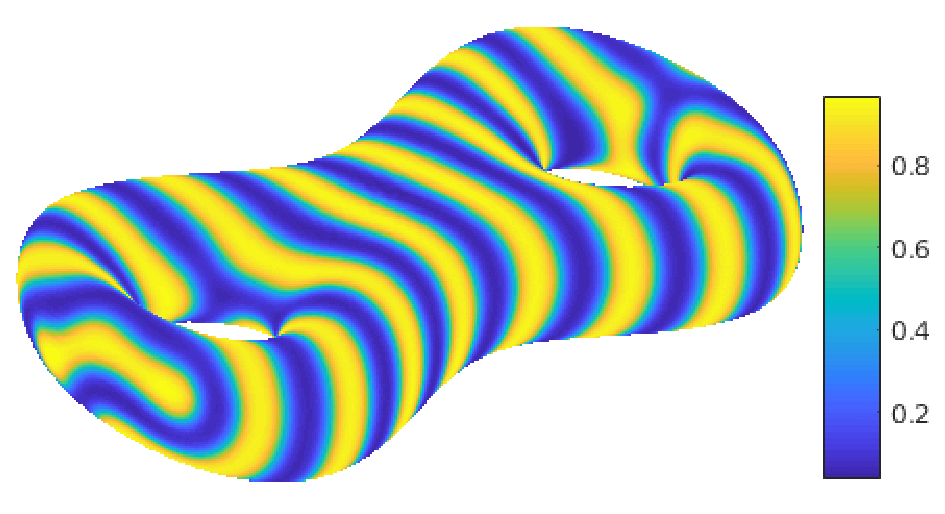}}
\hspace{0.5cm}
\subfigure[]{\includegraphics[scale=0.30]{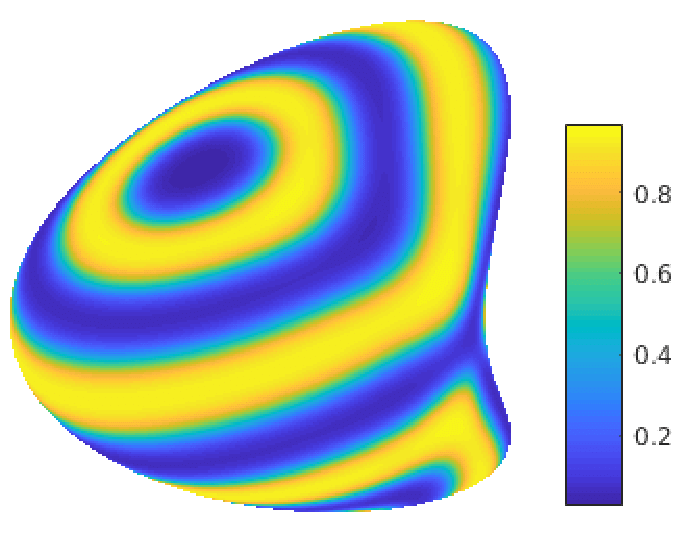}}
\hspace{0.5cm}
\subfigure[]{\includegraphics[scale=0.30]{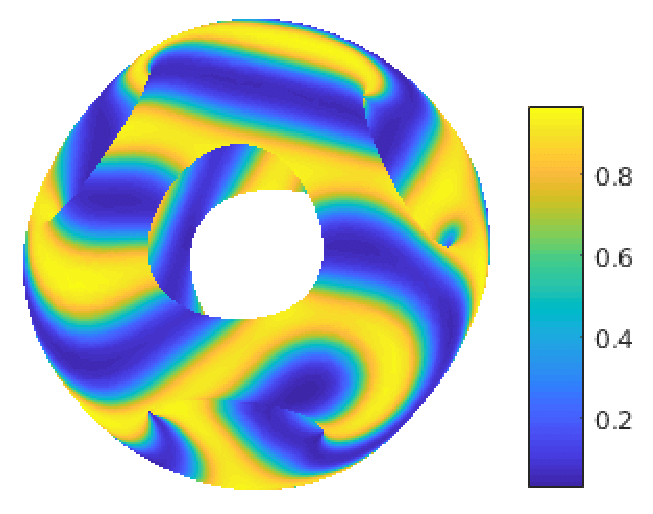}}
\caption{
    The spotted and striped structures of diblock copolymers on double torus,
    heart, and orthocircle surfaces. Yellow colors still correspond to large
    A-segment fractions.  For spotted phases, the parameter setting is $\chi
    N=25.0$ and $f=0.20$, while for striped patterns, $\chi N=15.0$ and
    $f=0.50$.
}
\label{fig:othersurface}
\end{figure}

As can be seen in Fig.\,\ref{fig:othersurface}, we have obtained spotted and
striped phases of diblock copolymers on these  closed surfaces. The initial
$\Gamma_0$ and optimal scale factor $\Gamma_{opt}$ of the three closed surfaces
are given in Tab.\,\ref{tab:general:radius}. The expressions of initial
surface $\calS_0$ have been given by
Eqns.\,\eqref{fig:mesh:doubletorus}-\eqref{fig:mesh:orthocircle}.  Obviously,
the scale factors of these three surfaces have been successfully optimized
dependent on the specific ordered structures.  For spotted structures
(Fig.\,\ref{fig:othersurface}(a)-(c)), the initial condition of the field functions
is chosen randomly. For symmetric systems, the striped phases
(Fig.\,\ref{fig:othersurface}(d)-(f)) are sensitive to the initial values. On
double torus and orthocircle surfaces, the morphologies of
Fig.\,\ref{fig:othersurface} (d) and (f) are obtained by setting the initial
value of $w_+$ and $w_-$ on each mesh node with coordinates $(x, y, z)$ as

\begin{align*}
    w_{-} = \chi N\sin(8x), ~~~ w_{+}  = 0.
\end{align*} 

For striped phase on heart surface as shown in Fig.\,\ref{fig:othersurface} (e),
we use the initial value as follows 
\begin{align*}
    w_{-} = \chi N\sin(8\theta), ~~~w_{+} = 0,
\end{align*}
where $\theta = \arctan(y/x)$, $\phi = \arctan(z/y)$, and $\theta, \phi \in [0, 2\pi]$.

\begin{table}[H]
	\caption{The initial $\Gamma_0$ and optimal scale factors $\Gamma_{opt}$ of 
	double torus, heart and orthocircle surfaces used in the
	surface simulations for diblock copolymers.}
  \label{tab:general:radius}
  \centering
\begin{tabular}{|c|c|c|c|c|c|c|}
 \hline
 & \multicolumn{2}{c}{Double torus} \vline & \multicolumn{2}{c}{Heart}
  \vline &\multicolumn{2}{c}{Orthocircle} \vline
 \\ \hline
    & Fig.\,\ref{fig:othersurface} (a) & Fig.\,\ref{fig:othersurface} (d) &
    Fig.\,\ref{fig:othersurface} (b) & Fig.\,\ref{fig:othersurface} (e) &
    Fig.\,\ref{fig:othersurface} (c) & Fig.\,\ref{fig:othersurface} (f)
 \\ \hline
    $\Gamma_0$ & 12 & 12 & 5 & 5 & 5 & 5 
    \\ \hline
    $\Gamma_{opt}$ &  12.308  & 14.161 & 5.022  & 5.403   & 5.143 & 4.939
 \\ \hline
\end{tabular}
\end{table}

For the asymmetric diblock copolymers, the authors in \cite{li2014self} have
found that the elongated spots of a spotted pattern usually
locate at the saddle points of a general curved
surface. In our simulations, this phenomenon does not appear.  The discrepancy
may be attributed to the discretization precision and the interaction strength.
In \cite{li2014self}, only $1912$ nodes were used in spotted phase
simulations.  However, our calculation uses, at least, $21560$ nodes (10782
elements) to discretize the surface which is enough to ensure the
discretization precision. Furthermore,  
the interaction parameter used here is $\chi N=25.0$ which is stronger than
that in \cite{li2014self} where $\chi N =13.0$.  The large interaction
parameter means strong segregation which contributes to the microscopic phase
separation.

\subsubsection{Parabolic surface }
\label{subsubsec:parabolic}

\begin{figure}[H]
\setlength{\captionmargin}{2pt}
\centering
\subfigure[]{\includegraphics[scale=0.35]{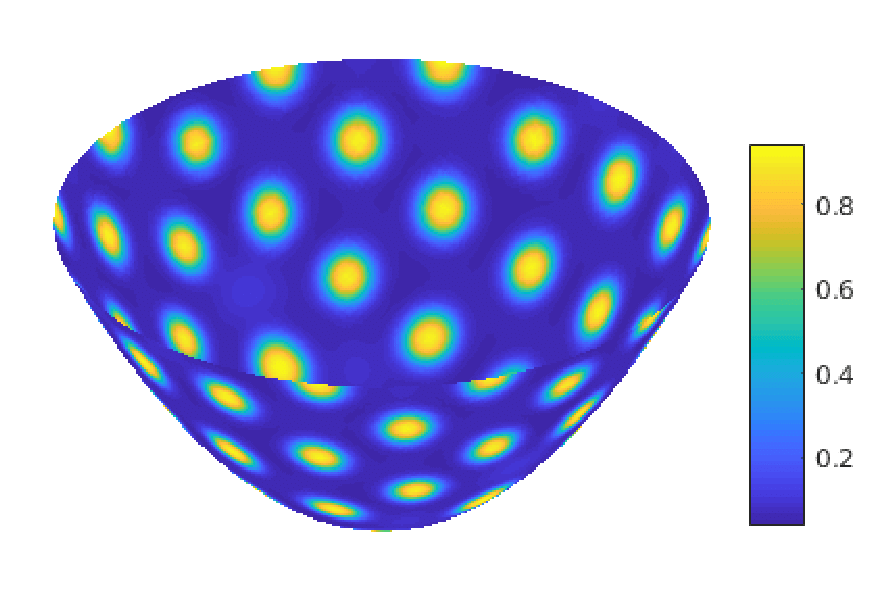}}
    \hspace{0.5cm}
\subfigure[]{\includegraphics[scale=0.35]{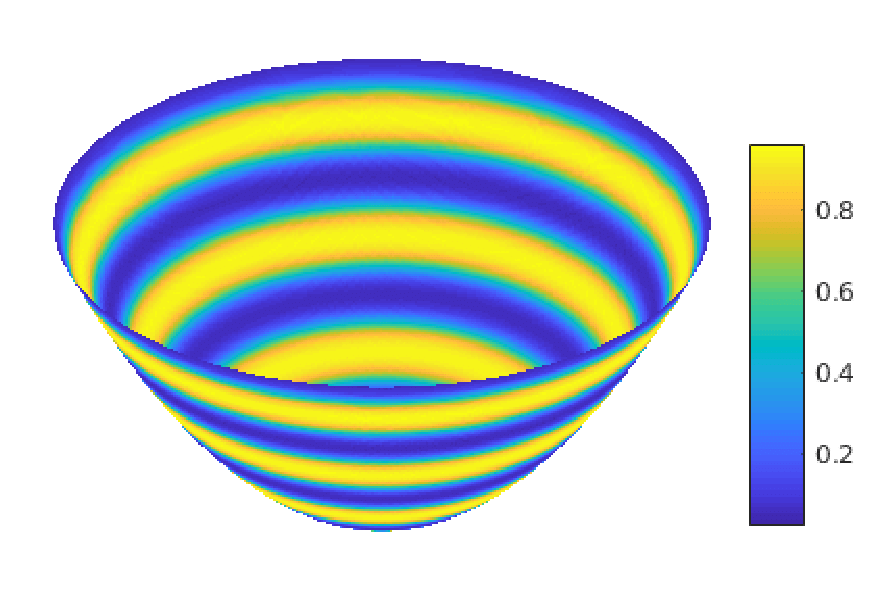}}
\caption{\label{fig:parabolic}
The ordered patterns of diblock copolymers on the parabolic surface
are obtained through linear surface finite element SCFT simulations. 
The homogeneous Neumann boundary condition is applied.
Yellow colors correspond to large A-segment fractions.
(a) Spotted phase when $\chi N = 25.0$, $f = 0.20$.
(b) Striped pattern when $\chi N = 15.0$, $f = 0.50$.
}
\end{figure}

\textcolor{blue}{ The above numerical experiments are all applied
on closed surfaces. In this subsection we apply the surface
finite element method to the SCFT model on a unclosed surface.
In particular, we solve SCFT equations on the parabolic surface.
For the unclosed surface, the boundary conditions should be
considered. In the finite element framework, 
Dirichlet, Neumann, mixed, or other boundary
conditions are all easily supported.
Here, we use the homogeneous Neumann boundary condition as an example to
test our method. Other boundary conditions can be also used in our
proposed numerical framework.} The initial values of $w_+$ and $w_-$ on
coordinates $(x, y, z)$ are given by $w_- = \chi N \sin(k z)$, $w_+=0$ both
for asymmetric and symmetric systems. $8515$ nodes and $16720$ elements are
used in the following simulations. When $k=5$, we obtain spotted ($\chi
N=25.0$, $f=0.20$) and striped ($\chi N=15.0$, $f=0.50$) phases, as shown
in the Fig.\,\ref{fig:parabolic}.   The initial scale factor
$\Gamma_0$ and
optimal $\Gamma_{opt}$ are given in Tab.\,\ref{tab:parabolic}. Results show
that our linear surface finite element method can efficiently capture the
standard self-assembled structures of block copolymers both in asymmetric
and symmetric cases on parabolic surface.  Meanwhile, the size of
the parabolic surface can be adaptively optimized during the SCFT
iteration.
\begin{table}[!htpb]
    \caption {The initial scale factor $\Gamma_0$ and optimal
	$\Gamma_{opt}$, in computing ordered patterns of diblock
	copolymers on parabolic surface.}
  \label{tab:parabolic}
  \centering
\begin{tabular}{|c|c|c|}
 \hline
    & Fig.\,\ref{fig:parabolic} (a) & Fig.\,\ref{fig:parabolic} (b) 
 \\ \hline
    $\Gamma_0$ & 10 & 10 
    \\ \hline
    $\Gamma_{opt}$ & 10.310 & 10.558
    \\ \hline
\end{tabular}
\end{table}

\section{Conclusion and Outlook}
\label{sec:conclusion}

In this paper, we proposed a linear surface finite element method
to solve the SCFT model and study the self-assembly behaviors of block copolymers on
general curved surfaces. 
At the same time, the surface size indeed
has been optimized during iteration procedure which can capture the
characteristic size of a given self-assembled structure, and give
a more accurate value of effective Hamiltonian. 
To demonstrate the ability of this algorithm, we
applied it to diblock copolymer systems on several distinct
curved surfaces, including 5 closed surfaces and an unclosed 
surface. Numerical results illustrate that our method
can successfully obtain the ordered structures of block
copolymers on these general curved surfaces. 

Block copolymers provide a perfect platform to study the self-assembly
behaviors in related physical, chemical and biology systems.  To study the
phase behaviors of block copolymers precisely, it still requires a high
accurate numerical method.  To ensure enough precision, we have to use many
nodes and elements to describe the ordered structures which costs much CPU
time.  In the future, we will further improve this computational approach in
many directions. The high order finite element method, non-uniform mesh, and
adaptive technique will be chosen to improve the precision and reduce
computational cost. Another interesting problem is to extend this
computational framework to more real-life polymeric systems, for example rigid
polymers.

\section*{Acknowledgements} This work is supported by National Science
Foundation of China (11771368, 11871413, 91430213, 91530321), and 
Project of Scientific Research Fund of Hunan Provincial Science
and Technology Department (2018WK4006).
H. Wei is partially
supported by  Hunan Provincial Civil-Military Integration Industrial
Development Project. K.~Jiang is partially supported by the research grand from
Hunan Science Foundation of China (2018JJ2376), and Youth Project Hunan
Provincial Education Department of China (Grant No. 16B257).

\appendix

\end{document}